\documentclass[journal,draftcls,onecolumn]{IEEEtran}
\usepackage{latexsym}
\usepackage{amsmath,amssymb,amsthm}
\usepackage{psfrag,array}
\usepackage{graphicx}
\usepackage{url}
\usepackage{fancyhdr}
\usepackage{color}

\usepackage{color,hyperref}
\definecolor{darkblue}{rgb}{0.0,0.0,0.3}
\hypersetup{colorlinks,breaklinks,
            linkcolor=darkblue,urlcolor=darkblue,
            anchorcolor=darkblue,citecolor=darkblue}

\renewcommand{\vec}[1]{\ensuremath{\boldsymbol{#1}}}
\newtheorem{lemma}{Lemma}
\newtheorem{theorem}{Theorem}

\newtheorem{proposition}{Proposition}

\newcommand{\stxt}[1]{\ensuremath{_{\text{#1}}}}
\newcolumntype{C}{>{$}c<{$}}
\newlength{\eqspace}
\newcommand{\defeq}{\stackrel{\textrm{def}}{=}}

\begin{document}


\title{Coupling Data Transmission for Multiple-Access Communications}

\author{Dmitri Truhachev and Christian Schlegel\footnote{This work was partially supported by NSERC Discovery Grant  and TELUS Corporation Canada and presented in part in ISIT 2013~\cite{isit13}.}}




\maketitle

\begin{abstract}
We consider a signaling format where the information to be communicated from one or multiple transmitters to a receiver is modulated via a superposition of independent data streams. Each data stream is formed by error-correction encoding, constellation mapping, replication and permutation of symbols, and application of signature sequences. The relations between the data bits and modulation symbols transmitted over the channel can be represented by a sparse graph. In the case where the modulated data streams are transmitted with time offsets the receiver observes spatial coupling of the individual graphs into a graph chain enabling efficient demodulation/decoding. We prove that a two-stage demodulation/decoding method, in which iterative demodulation based on symbol estimation and interference cancellation is followed by parallel error correction decoding, achieves capacity on the additive white Gaussian noise (AWGN) channel asymptotically. We compare the performance of the two-stage receiver to the receiver which utilizes hard feedback between the error-correction encoders and the iterative demodulator.
\end{abstract}

\section{Introduction}

Recently, the technique of spatial graph coupling applied to iterative processing on graphs has attracted significant interest in broad areas of communications. The method was first introduced to construct low-density parity-check convolutional codes (LDPCCCs)~\cite{fz99} that exhibit the so-called {\em threshold saturation} behavior~\cite{lscz10}\cite{kru11} which occurs when the limit (threshold) of suboptimal iterative decoding of LDPCCCs asymptotically achieves the optimum maximum a posteriori probability (MAP) decoding threshold~\cite{kru11} of underlying LDPC block codes with the same structure~\cite{kru12u}. The idea of constructing graph structures from connected identical copies of a single graph has since been applied to compressed sensing~\cite{kp10}\cite{dmj11}, image recognition, quantum coding~\cite{hkis11}, and other fields.

A number of applications of spatial graph coupling to communications over multi-user channels have since been developed. Yedla et. al. studied threshold saturation for a two-user Gaussian multiple access channel (MAC) \cite{gmaccoup11} while Kudekar and Kasai studied it for the binary erasure MAC~\cite{bemac}. In terms of multi-user signaling formats spatially-coupled partitioned-spreading code-division multiple-access (PS-CDMA) was studied in~\cite{st11,TruComLet12}. It was proven that in a noiseless regime the coupled system can achieve arbitrary multi-user loads with the two-stage demodulation/decoding method contrary to the uncoupled PS-CDMA~\cite{TruSchKrzIT09} or dense CDMA in general~\cite{tan02}.  
Sparse CDMA, extensively studied by Guo and Wang~\cite{sparseCDMA08} in the uncoupled regime, was extended via spatial coupling in~\cite{ttk11,TakeuchiTK12} where significant threshold improvements over uncoupled CDMA were observed.

In this work we focus on a generalized multi-user communication signaling format which lands itself to efficient spatial coupling and iterative demodulation. Each user's transmitter chain includes common elements such as error-correction encoding, higher-order constellation mapping, symbol repetition and permutation, and signature sequences. Its complexity and rate can be flexibly adjusted depending on the communication scenario. The system can be operated and studied in both coupled and uncoupled (block) regimes. For specific parameter settings some classes of coupled generalized modulation~\cite{GM10}, PS-CDMA~\cite{st11}, or IDMA introduced by Ping et. at.~\cite{PiLiuWuLeu06,LiuTonPin06,HoeSchFri08} can be derived as particular cases. 
The iterative demodulation performed at the receiver is a sequence of symbol estimation and interference cancellation iterations and can be seen as a message-passing process on the system graph. 

In this paper we investigate spectral efficiency achievable by iterative demodulation/decoding in block and coupled regimes and the resulting gap to the channel capacity. In particular we focus on the two-stage demodulation/decoding schedule where feedback between the error-correction decoders and the demodulator is absent.  The two-stage schedule allows for a simplified system design and improved decoding latency compared to either sequential peeling schedules or turbo demodulation/decoding schedules based on soft information exchange between the demodulator and the error-correction decoders.  The main contributions of our work can be summarized as follows.

\subsection{Main Contributions}

\begin{itemize}
\item We rigorously derive a coupled recursion characterizing the evolution of noise-and-interference powers throughout demodulation iterations for both block and coupled versions of the system. We show that for any given number of demodulation iterations there exists a system graph for which the iterative demodulation operates as a message passing on a set of trees. We then quantify the achievable data rate as a function of the signal-to-noise ratio (SNR), constellation symbol alphabet, and the sets of finite system parameters: the number of interfering data streams, repetition factor, and symbol constellation.
\item For the case of binary symbols we derive an upper bound on the gap between the achievable rate and the channel capacity. The gap is a function of the SNR and the finite set of system parameters. We show that the asymptotic gap  is inverse proportional to the capacity itself and vanishes as the system's SNR increases.
\item We derive an expression for the asymptotic scaling of the gap between the achievable sum-rate and the channel capacity.
\item We study  an alternative demodulation/decoding schedule were hard decision feedback from the error-correction decoders to the demodulator is allowed. We derive a bound on the gap to capacity and compare the achievable rate with that of the two-stage schedule. Finally, we prove that in case of hard decision feedback the gap to capacity vanishes for any fixed SNR as the system load increases.
\end{itemize}

\subsection{Related Work}


A related multi-user communications problem is the problem of maximizing the number of users simultaneously communicating over the MAC channel or achieving high multi-user efficiency~\cite{ver86} measuring the impact of residual post/detection  multi-user interference on a user performance. A number of papers have been dedicated to study asymptotics of joint and individual multiuser detection for classic CDMA. Tse and Verdu investigated the asymptotic performance of joint MAP decoding of random CDMA, and proved that the multiuser efficiency approaches unity as the SNR goes to infinity~\cite{tsever00}. 
Later Tanaka~\cite{tan02}, M\"uller and Gerstacker~\cite{mulger04}, and Guo and Verdu~\cite{guover05} studied the performance of joint and individual multi-user detection of CDMA via the non-rigorous replica method, and derived the limits for the rates of coded random CDMA with joint and separate demodulation/decoding in terms of multi-user efficiency. In a paper of Takeuchi, Tanaka, and Kawabata~\cite{ttk12perf}, developed in parallel to ours~\cite{isit13}, coupled sparse CDMA was studied in the asymptotic regime where the number of users, spreading gain and the system size were taken to infinity, while  the multi-user load was kept constant. Based on a continuous approximation of the respective coupled recursion for the case of binary symbols the authors showed saturation of the iterative detection threshold of sparse CDMA to the MAP threshold of dense CDMA via spatial graph coupling. 

The focus of our paper is on rigorously deriving and numerically demonstrating the rates achievable by the block and coupled sparse multi-user system we consider, where we quantify the resulting gap to capacity explicitly as a function of SNR rather than multi-user efficiency.  We study a generalized system in terms of symbol alphabets and derive the achievable rate for finite parameter settings in terms of the number of data streams, repetition factor, symbol constellation alphabet etc. Our approach allows us to study the system performance for various explicitly defined decoding schedules such as the demodulation/decoding schedule with hard feedback, numbers of iterations, and receiver signal-processing options. Our framework leads to graphical interpretations of the achievable rates and the gap to capacity in a form similar to the  threshold saturation theorem for LDPCCC in BEC~\cite{kru11}.

%

It's worth mentioning that the capacity of the multiple-access channel can be achieved using peeling decoding, the fact that follows directly from the chain rule of the mutual information. Various types of the peeling or sequential cancellation decoders has been studied extensively for a variety of systems~\cite{ih77,DuaRimUrb97,MaPing,Cro10}. 
The focus of this work, however, is on iterative demodulation using the low-latency two-stage demodulation/decoding schedule and, in particularly on the role of spatial graph coupling in approaching channel capacity.

Interleaved-division modulation (IDM), a modulation format utilized in point-to-point transmission proposed by Hoeher et. al.~\cite{HoeSchFri08,HoeWo11}, has common elements with the multi-user format we consider here, such as superposition of data streams and bit interleaving, which is also a feature of bit-interleaved coded modulations in general~\cite{ctb98}. Contrary to the reception techniques geared towards separation of symbol-synchronous data streams studied by Hoeher and Wo, we investigate cancellation-based low-complexity iterative interference removal which is agnostic to data stream synchronism and power fluctuations and is specially well-suited for coupling.

Another related signaling format are the sparse superposition codes studied by Baron and Joseph~\cite{spark10}, later shown to achieve capacity in the block unequal power regime~\cite{spark15}. Sparse superposition codes are based on superimposing data streams based on Gaussian matrices, the technique which is closely related to compressed sensing reconstruction. A coupled version of sparse superposition codes was recently studied in~\cite{Barbier16} where it was shown that, given the coupled recursion describing the state evolution of the approximate message-passing (AMP) decoder, the codes achieve channel capacity asymptotically. 
The AMP decoder of coupled superposition coding does not lend itself to a rigorous derivation of the coupled recursion, however.


\subsection{Structure of the Paper}

System model, transmission format and receiver processing are described in Section~\ref{sec:sys}. Section~\ref{sec:main} presents our performance analysis, derivation of the main coupled recursions and formulation of the main results. Numerical results, discussion and relation to optimal MAP demodulation/decoding are given in Section~\ref{sec:sims}. 
Finally Section~\ref{sec:conc} concludes the paper.

\section{System Model}
\label{sec:sys}

We consider a communication scenario in which one or more transmitting terminals communicate to a single receiver. The signal observed at the receiver is formed by a superposition of $L$ independently modulated data streams that originate at the terminals and then superimpose at the receiver. 

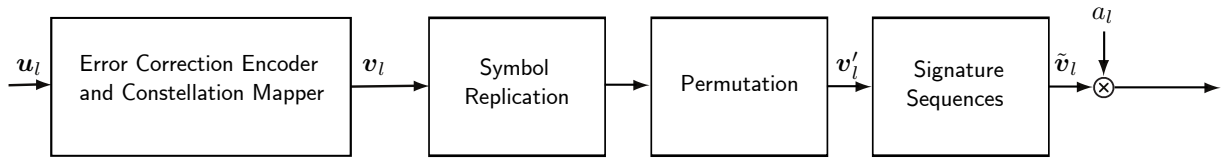
\begin{figure}[t]
\setlength{\unitlength}{1mm}
   \begin{picture}(155,32)
   \put(145,23.5){$a_l$}
   \put(2,16.5){$\vec{u}_l$}
   \put(140,16.5){$\tilde{\vec{v}}_l$}
   \put(111,16.5){$\boldsymbol{v}'_l$}
    \put(48,16.5){$\vec{v}_l$}
   \put(-2,0){\includegraphics{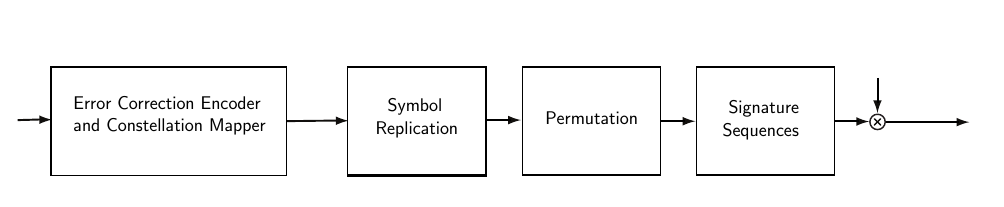}
}
\end{picture}
\caption{Generation of the $l$th data stream at the transmitter side.}
\label{Fig:Sys}
\end{figure}

The processes of generating data stream $l$, $l=1,2,\cdots,L$ is depicted in Fig.~\ref{Fig:Sys}. First a binary information sequence $\boldsymbol{u}_l = u_{1,l},u_{2,l},u_{3,l},\cdots, u_{K,l}$ is encoded by a binary error correction code of rate $R$ and mapped to a sequence of signal constellation points denoted by $\vec{v}_l= v_{1,l},v_{2,l},v_{3,l},\cdots,v_{N,l}$, where $v_{n,l} \in \mathcal{A}_l$. By $\mathcal{A}_l$ we denote a set of signal constellation symbols (which can be a PAM constellation, for example). At the next step, each symbol of the sequence $\boldsymbol{v}_l$ is replicated $M_1$ times. After the replication the resulting symbols are permuted using a binary permutation matrix ${\bf P}_l$ of size $NM_1 \times NM_1$ specific to the data stream $l$, $l=1,2,\cdots,L$.  

Finally each symbol ${v}'_{i,l}$, $i=1,2,\cdots,NM_1$, $l=1,2,\cdots,L$ of the permuted sequence is multiplied by a signature sequence $\boldsymbol{s}_{i,l}$ of length $M_2$. We consider pseudo-random energy normalized binary signature sequences, however, Gaussian or other sequences can also be used. The symbols of the signature sequences satisfy 
\begin{equation}
\mathbb{E}s_{n_1,l_1,m_1}s_{n_2,l_2,m_2} =  
\begin{cases} 
      1 & \textrm{if } (n_1,l_1,m_1)=(n_2,l_2,m_2) \\        0 & \textrm{otherwise}\ .
 \end{cases}
\end{equation}
The resulting sequence $\tilde{\boldsymbol{v}}_l = \tilde{v}_{1,l},\tilde{v}_{2,l},\cdots,\tilde{v}_{MN,l}$ is multiplied by an amplitude $a_l$ and transmitted over the channel. The overall symbol repetition factor is given by $M = M_1 M_2$. 

Hence, generation of data streams is done in two steps. The first step encompasses individual error-correction encoding and constellation mapping. The second step is the preparation of the data for multi-user processing at the receiver and includes replication, permutation, and application of the signature sequences.

\subsection{Block System}
\label{subsec:uncoup}

\begin{figure}[h]
\setlength{\unitlength}{1mm}
   \begin{picture}(155,90)
   \put(-2,0){\includegraphics{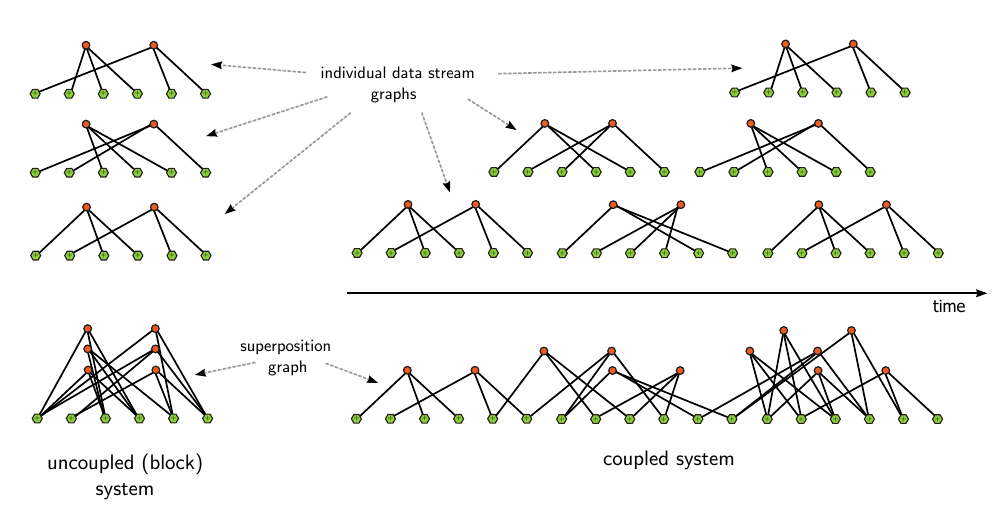}}
\end{picture}
\caption{Graph representation of the block and spatially coupled systems.}
\label{Fig:graphs}
\end{figure}

The {\em block system} scenario is a communication scenario in which all data streams are transmitted over the channel simultaneously. Assuming a real-valued Gaussian noise channel, the received signal is given by 
\[
\boldsymbol{y} = \sum_{l=1}^L a_l \tilde{\boldsymbol{v}}_l + \boldsymbol{n}\ 
\]
where $\vec{n}$ is a vector of the iid Gaussian noise samples with zero mean and variance $\sigma^2$. The block system can be represented by the graph shown on the left side of Fig.~\ref{Fig:graphs}. The top part of the figure demonstrates three individual data stream graphs while the resulting graph observed at the receiver is shown at the bottom. Variable nodes representing symbols $v_{j,l}$ are depicted by circles and channel nodes representing received values $y_t$ are depicted by hexagons. 

\subsection{Coupled System}
\label{subsec:coup}

A spatially graph-coupled system for the presented modulation format arises naturally  when the modulated data streams are transmitted over the channel with time offsets. 
This scenario is illustrated in the right side of Fig.~\ref{Fig:graphs}. The graphs representing transmitted data streams couple through the channel nodes that correspond to the times at which the data stream symbols are transmitted over the common channel. Contrary to the block system situation, where the data streams superimpose in groups of $L$ and each group is represented by a graph at the receiver, the coupled system is represented by a single large graph also called the coupled graph chain. This draws parallels to coupled graph chains  appearing in other applications such as SC-LDPC codes~\cite{fz99}. We will prove that iterative demodulation and decoding on this graph chain will give us substantially improved achievable rates.

The hexagonal channel nodes represent received symbols $y_t$, $t=1,2,3,\cdots$. The symbols $\tilde{v}_{j,l}$ transmitted at the same time $t$ are connected to the same channel node. For the sake of graph representation we assume that the transmission is ``symbol-synchronous''. It is important to note, however, that the receiver processing which is based on symbol estimation and interference cancellation is not based on this assumption nor is the iterative demodulation analysis. 

We denote the time offset (delay) of data stream $l$ by $\tau_l$, $l=1,2,\cdots,\hat{L}$ where $\hat{L}$ is the total number of streams transmitted. Then the components of the received signal vector $\vec{y}$ are given by
\[
y_t = \sum_{l=1}^{\hat{L}} a_l \tilde{v}_{t-\tau_l,l}+ n_t\ ;  \quad \quad t=1,2,3 \cdots.
\]

In order to study the performance of the coupled system we consider a regularized delay structure defined as follows. We introduce $W$, the coupling parameter and assume that each transmitted data stream consists of $2W+1$ subsections of length $N\stxt{w}=MN/(2W+1)$. In addition the data streams are transmitted in groups of size $L\stxt{W}=L/(2W+1)$ streams. At time $t=1$ the first group of data streams is transmitted, i.e., the first $L\stxt{W}$ time delays are given by $\tau_1= \tau_2 =\cdots =\tau_{L\stxt{W}}= 0$. At time $N\stxt{W}+1$ the second group of  $L\stxt{W}$  data streams is transmitted with the corresponding time delays given by $\tau_{L\stxt{W}+1}=\tau_{L\stxt{W}+2} =\cdots =\tau_{2L\stxt{W}}= N\stxt{W}$ and so on. At every time $kN\stxt{W}+1$, $k=0,1,2,\cdots$  exactly $L\stxt{W}$ data streams are transmitted. In the remainder of the paper we use the terminology  ``coupled system'' to address the system with regularized delays. The regularized delay structure is assumed for analysis purposes and to demonstrate the capacity-achieving properties of the coupled system and is not required for receiver signal processing. Poisson delays produce results close to those with regularized delays, however, the study of delay distributions and their impact is beyond the scope of this paper.

While the degrees of nodes in the coupled graph chain are predetermined, the connections between the nodes (edges) are not. We can define a graph ensemble describing the coupled system by considering random permutation matrices ${\bf P}_l$ in the generation of the individual data streams. This representation is given in Appendix~\ref{app:cycle_free} along with a matrix representation of the coupled chain. In addition, signature sequences are also chosen randomly.  The right side of Fig.~\ref{Fig:graphs} illustrates an example in which six data streams transmitted with delays $\tau_1=1$, $\tau_2=5$, $\tau_3=7$, $\tau_4=11$, $\tau_5=12$, $\tau_6=13$. 

\subsection{System Load and Rate}
\label{subsec:load}

We define the load $\alpha$ of the block system by taking the ratio of the number of simultaneously transmitted data streams $L$ to the repetition factor $M$, $\alpha=L/M$. Assume that all data streams have the same power normalized to $1/L$ and constellation alphabets $\mathcal{A}_l$, $l=1,2,3,\cdots$ that consists of $2^B$ signal points, where $B$ is the modulation index. The total system SNR measured as a ratio of the total transmit power over the noise power then equals $1/\sigma^2$ while the total transmit data rate equals $RBL/M=\alpha BR$ information bits per channel use. The signal-to-noise ratio per information bit equals $E\stxt{b}/N_0=1/(2 \alpha B R \sigma^2)$.


In the coupled case the system load in terms of the number of data streams per repetition factor $M$ for the first $NM$ time instances $t=1,2,\cdots,NM$ is smaller than $\alpha$. Assuming an infinite number of data streams in the system, $\hat{L}=\infty$ and constant system load $\alpha$ for $t>N$  we obtain the same  rate and SNR as in the block  case. The initial reduced load leads to an initial rate loss that vanishes as $\hat{L} \rightarrow \infty$. This is typical for all coupled systems including the one we consider. 


\subsection{Iterative Demodulation}

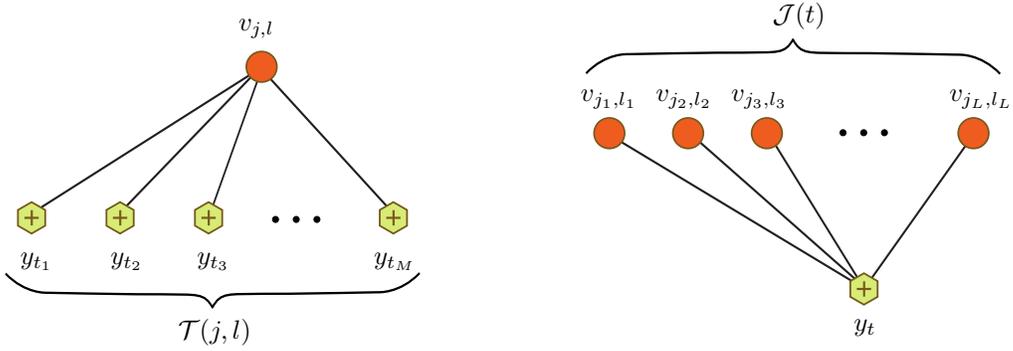
\begin{figure}[t]
\setlength{\unitlength}{1mm}
   \begin{picture}(155,50)
   \put(35,3){$\mathcal{T}(j,l)$}
   \put(43,44){$v_{j,l}$}
   \put(14,12.9){$y_{t_1}$}
   \put(26,12.9){$y_{t_2}$}
   \put(37.5,12.9){$y_{t_3}$}
   \put(61,12.9){$y_{t_M}$}
   \put(114,45){$\mathcal{J}(t)$}
   \put(88.5,34.5){$v_{j_1,l_1}$}   
   \put(98.5,34.5){$v_{j_2,l_2}$}   
   \put(108.5,34.5){$v_{j_3,l_3}$}
   \put(137.5,34.5){$v_{j_L,l_L}$}      
   \put(124.8,4){$y_t$}
   \put(0,0){\includegraphics{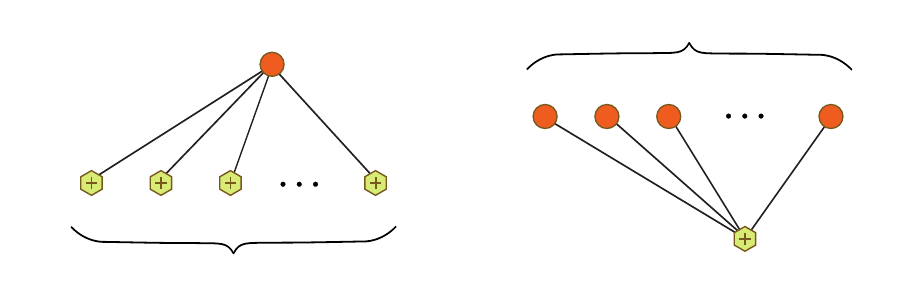}
}
\end{picture}
\caption{The relation between the variable and channel nodes and the sets of indices involved. }
\label{Fig:VarGraph}
\end{figure}

The received signal vector $\boldsymbol{y}=(y_1,y_2,y_3,\cdots)$ contains $M=M_1M_2$ replicas of each transmitted symbol $v_{j,l}$, $j=1,2,\cdots$, $l=1,2,\cdots,L$. Let $\mathcal{T}(j,l)=\{ t_1,t_2, \cdots,t_M \}$ denote the set of time indices $t$ such that the received vector component $y_t$, $t \in \mathcal{J}(t)$ contains $v_{j,l}$. Let's denote the corresponding symbols of the signature sequences that multiply $v_{j,l}$ during the modulation process by $s^{(t)}_{j,l}$, where $t \in \mathcal{T}(j,l)$. Moreover, by $\mathcal{J}(t)$ we denote a set of all index pairs $(j,l)$ such that $v_{j,l}$ is included in $y_t$ (see Fig.~\ref{Fig:VarGraph}). The cardinality of the set $\mathcal{T}(j,l)$ is always equal to $M$. The cardinality of $\mathcal{J}(t)$ equals $L$ for a block system and varies depending on $t$ for the coupled system (see Section~\ref{subsec:load}). 

Demodulation iteration $i$ starts with the interference cancellation step computing signals $y^{(i)}_{t\rightarrow (j,l)}$ for $j=1,2,\cdots,N$, $l=1,2,\cdots,L$, $t \in \mathcal{T}(j,l)$
\begin{align}
{y}^{(i)}_{t\rightarrow (j,l)} &= {y}_t - \mathop{\sum_{(j',l') \in \mathcal{J}(t)}}_{\textrm{s.t. } (j',l') \neq (j,l)} a_{l'} s^{(t)}_{j',l'} \hat{v}^{(i-1)}_{(j',l') \rightarrow t},   
\label{eq:ici} \\
 &=  a_l s^{(t)}_{j,l} v_{j,l} + \xi_{j,l,t} + n_t,
\end{align}
where 
\begin{equation}
\xi_{j,l,t} \defeq \mathop{\sum_{(j',l') \in \mathcal{J}(t)}}_{\textrm{s.t. } (j',l') \neq (j,l)} a_{l'} s^{(t)}_{j',l'} \left(v_{j',l'}-\hat{v}^{(i-1)}_{(j',l') \rightarrow t}\right)
\label{eq:ici2}
\end{equation}
is the residual interference component and  the estimates $\hat{v}^{(i-1)}_{(j',l') \rightarrow t}$ have been generated for all data symbols $v_{j',l'}$, $j'=1,2,\cdots$, $l'=1,2,\cdots,L$ at the previous $(i-1)$th iteration.\footnote{At the first iteration $i=1$ we use $\hat{v}^{(0)}_{(j',l')\rightarrow t}=0$ (no interference cancellation is attempted).} 

For each $\tau \in \mathcal{T}(j,l)$ we form a vector $\vec{y}^{(i)}_{j,l,\tau}$ out of estimates $\vec{y}^{(i)}_{(j,l) \rightarrow t}$ where $t \neq \tau, t \in \mathcal{T}(j,l)$. Since each signal $\vec{y}^{(i)}_{j,l,\tau}$ contains $v_{j,l}$ and some remaining interference we can write
\begin{equation}
\vec{y}^{(i)}_{j,l,\tau} = \vec{h}_\tau v_{j,l} + \vec{\xi}_{j,l,\tau} \quad \quad  \tau \in \mathcal{T}(j,l) 
\label{eq:xijlt}
\end{equation}
where the vector $\vec{h}_\tau$  consists of elements $a_ls^{(t)}$,  $t \neq \tau, t \in \mathcal{T}(j,l)$ and $\vec{\xi}_{j,l,\tau}$ is the noise-and-interference vector with components given in (\ref{eq:ici2}). Let us denote the covariance matrix of the noise-and-interference vector $\vec{\xi}_{j,l,\tau}$ by  $\vec{R}_{j,l,\tau}$. We now perform minimum mean-squared error (MMSE) filtering of $\vec{y}^{(i)}_{j,l,\tau}$ to form an SNR-optimal linear estimate of $v_{j,l}$, given by
\begin{equation}
z_{j,l,\tau} = \vec{w}^\textrm{T}_{j,l} \vec{y}^{(i)}_{j,l,\tau} 
\label{eq:zestimate}
\end{equation}
where
\[
\vec{w}^\textrm{T}_{j,l,\tau} = (\vec{I}+\vec{h}_\tau^*\vec{R}_{j,l,\tau}^{-1}\vec{h}_\tau)^{-1} \vec{h}_\tau^*\vec{R}_{j,l,\tau}^{-1} \quad \quad  t \in \mathcal{T}(j,l)
\]
minimizes $||\vec{w}^\textrm{T}_{j,l,\tau} \vec{y}^{(i)}_{j,l,\tau}-v_{j,l}||^2 $. 


Since we know that $v_{j,l}$ belongs to the symbol alphabet $\mathcal{A}_l$ we can form a conditional expectation estimate 
\begin{equation}
\hat{v}^{(i)}_{(j,l) \rightarrow \tau} =  \mathbb{E} (v_{j,l}|z_{j,l,\tau}) \quad \quad  \tau \in \mathcal{T}(j,l)\ .
\label{eq:vestimate}
\end{equation}
For the case of equiprobable binary symbols $v_{j,l} \in \{1,-1\}$ (2-PAM modulation)  $\hat{v}^{(i)}_{(j,l) \rightarrow \tau}$ takes the form $\hat{v}^{(i)}_{(j,l)\rightarrow \tau} = \tanh(z_{j,l,\tau} \gamma_{j,l,\tau})$, where $\gamma_{j,l,\tau}$ is the SNR of the estimate $z_{j,l,\tau}$. We will compute this SNR in the next section.

This rules correspond to message passing on a system's graph. Each variable node $(j,l)$ passes $M$ estimates $\hat{v}^{(i)}_{(j,l) \rightarrow t}$ to the connected channel nodes $t \in \mathcal{T}(j,l)$ while each channel node passes $z^{(i)}_{j,l,t}$ where $ (j',l') \in \mathcal{J}(t)$ to the connected variable nodes.  

\subsection{Demodulation and Decoding Schedules}
\label{sec:decodingschedules}

The demodulation and error-correction decoding at the receiver can be scheduled in a number of ways. We consider two demodulation/decoding schedules which are of relatively low complexity. An alternative schedule of significantly higher complexity is the ``turbo-schedule'' with soft feedback between the error-correction decoders and the iterative demodulator.

\vspace{3mm}
\noindent{\it Two-Stage Schedule}
\vspace{3mm}

At the first stage $I$ iterative demodulation iterations are performed as described above. The second stage comprises decoding of the forward error correction codes used to encode the information vectors $\vec{u}_l$, $l=1,2,\cdots,\hat{L}$. 

\vspace{15mm}
\noindent{\it Hard Decoding Feedback and Cancellation Schedule}
\vspace{3mm}

Demodulation iterations are performed until one or more data stream's SNR rise above the decoding threshold of the error-correction code. Once this happens the code is decoded and the decoded bits are given as hard feedback to the demodulator. The feedback bits are then utilized to modulate the respective data streams that are subtracted (cancelled) from the received signal. The demodulation process continues for the remaining undecoded data streams until one of these would have a sufficiently high SNR to be decoded, and so on.



\section{Performance Analysis}
\label{sec:main}





At every iteration the demodulator attempts to reduce the amount of inter-stream interference. We take the approach to track the evolution of the signal-to-noise and interference ratio (SINR) throughout the decoding iterations~\cite{BurSchShiKry2004,LiuTonPin06,allert06}. The central part of the performance analysis is dedicated to the derivation of a  noise-and-interference power evolution equation to track the evolution of the noise-and-interference power throughout the demodulation iterations. 

We start with an assessment of the performance of the block system for which the evolution of the noise-and-interference power is independent of the time index $t$. After that we derive a characteristic equation that determines when the iterative demodulation process for the block system converges to a nearly interference-free state.

We then proceed with deriving the noise-and-interference power evolution for the coupled system and make a connection to the characteristic equation of the block system and its fixed points. Based on this connection we obtain the expressions for the achievable rates and the gap to capacity. Contrary to related work we make our derivation for  finite parameter values of $M,L,B$ and then derive the asymptotically achievable rates as particular cases.

\subsection{Block System}
\label{subsec:IIIA}

We focus on the case of equal power data streams where the amplitudes are given by $a_l=a$, $l=1,2,\cdots,L$. Without loss of generality we normalize data stream powers so that $a=\sqrt{1/L}$ and vary the noise power $\sigma^2$. In this case the total system SNR per receive symbol equals $1/\sigma^2$ (see Section~\ref{subsec:load}). We then proceed with considering the evolution of the noise-and-interference power (\ref{eq:ici2}).

Let us denote the noise-and-interference power at iteration $i$ by 
\begin{align}
x_i &\defeq \mathbb{E}\left(\mathop{\sum_{(j',l') \in \mathcal{J}(t)}}_{\textrm{s.t. } (j',l')\neq (j,l)} \sqrt{\frac{1}{L}} s^{(t)}_{j',l'}(v_{j',l'} - \hat{v}_{(j',l') \rightarrow t}^{(i-1)}) + n_{\tau}\right)^2
 = \mathop{\sum_{(j',l') \in \mathcal{J}(t)}}_{\textrm{s.t. } (j',l')\neq (j,l)} \frac{1}{L}\mathbb{E}\left(v_{j',l'} - \hat{v}_{(j',l') \rightarrow t}^{(i-1)}\right)^2 + \sigma^2 \label{eq:xidef}\\
 &=\frac{1}{L}(L-1)\mu_{i-1}+\sigma^2
\end{align}
where the mean-square error (MSE) of the symbols at iteration $i-1$ is given by $\mu_{i-1}$, i.e.
\begin{equation}
\mu_{i-1} \defeq \mathbb{E}\left(v_{j',l'} - \hat{v}_{(j',l') \rightarrow t}^{(i-1)}\right)^2 =  \mathbb{E}(v_{j',l'} - \mathbb{E} (v_{j,l}|z_{j,l,t}))^2\ .
 \label{eq:muidef}
\end{equation}
The expectation is taken over the respective channel node index $t$, variable node index $(j',l')$, and the system realizations.
When breaking the variance of the noise-and-interference power in (\ref{eq:ici2}) into the sum of variances of the components $v_{j',l'} - \hat{v}_{(j',l') \rightarrow t}^{(i-1)}$ as in (\ref{eq:xidef}) we assume that these are independent. This can be guaranteed by making sure that the system graph has large girth, a property not required for practical implementation but important for an exact analysis.

Coming back to the graph representation of the system, if we connect the variable node $v_{j,l}$ to the channel nodes $y_t$ where $t \in \mathcal{T}(j,l)$ and then connect each channel node $y_t$ to $v_{j',l'}$ such that $v_{j,l} \in \mathcal{J}(t)$ we obtain a sub-graph describing one demodulation iteration for symbol $v_{j,l}$. If we expand it further by connecting the nodes $v_{j',l'}$ to their channel nodes $y_\tau \in \mathcal{T}(j,l)$ and so on we can obtain a sub-graph describing $i$ demodulation iterations for symbol $v_{j,l}$. If the girth of the entire system's graph is at least $4i+1$ any demodulation sub-graph for any symbols $v_{j,l}$ will not contain repeated nodes and is, therefore, a tree. This description and conditions closely resemble those for LDPC block~\cite{Gal06} and convolutional code analysis~\cite{isit2001}. The following lemma elaborates on this connection and proves that a system graph with arbitrary large girth can be constructed for both block and coupled systems if the data stream length is chosen large enough.

\begin{lemma}
For any number of iterations $i$, and parameters $L,M,B,W$, both coupled and block systems can be constructed such that the corresponding graph has girth larger than $4i$.
\label{lem:cycle_free}
\end{lemma}
\begin{proof}
See Appendix~\ref{app:cycle_free}.
\end{proof}

The next step is to express the MSE $\mu_{i-1}$ in terms of the noise-and-interference power $x_{i-1}$ at iteration $i-1$. The Central Limit Theorem implies that the noise-and-interference vector $\vec{\xi}_{j,l,t}$ in (\ref{eq:xijlt}) converges to a Gaussian random vector with independent zero-mean components and covariance matrix $\vec{R}_{j,l,t} = \textrm{diag}(\sigma_{t_1}^2,\sigma_{t_2}^2,\cdots,\sigma_{t_M}^2)$ ($t$ excluded) as $L$ increases, where $\sigma_{\tau}^2$ denotes the variance of the individual vector components $\xi_{j,l,t}$, $\tau \in \mathcal{T}(j,l), \tau \neq t$. Typical values of $L$ are around $100$ or higher indicating that the Gaussian approximation is very accurate. 
 The Lindeberg condition is satisfied for both coupled and block equal power systems. The resulting SNR of $z_{j,l,t}$ in (\ref{eq:vestimate}) equals
\begin{equation}
\gamma_{j,l,t} = \vec{h}_t^*\vec{R}_{j,l,t}^{-1}\vec{h}_t = \frac{1}{L} \sum_{\tau \in \mathcal{T}(j,l) } \frac{1}{\sigma_\tau^2}\ .
\label{eq:gammaestimate}
\end{equation}
The MSE expression (\ref{eq:muidef}) can be computed as MSE of symbol estimation in additive Gaussian noise and is a function of the symbol constellation and the SNR. We denote it by
\begin{equation}
\mu_{i-1} = g\stxt{mse}(\gamma_{j,l})\ .
\label{eq:gmse}
\end{equation}
The MSE for the binary symbol alphabet case takes the form 
\begin{equation}
g\stxt{2}(\gamma) = \mathbb{E}\left[\left(1-\tanh\left(\gamma+\xi\sqrt{\gamma}\right)\right)^2\right]\ ; \quad\xi \sim \mathcal{N}(0,1)\ .
\label{eq:gmse}
\end{equation}
A series expansion and a number of bounds and approximations for (\ref{eq:gmse}) are derived in~\cite{BurSchShiKry2004}. 


In the case of the block system all noise-and-interference components (\ref{eq:ici2}) have the same power and, therefore, $\sigma_{\tau}^2=x_{i-1}$, $\tau \in \mathcal{T}(j,l) $ for all $k$ forming the diagonal covariance matrix $\vec{R}_{j,l}=\frac{1}{x_{i-1}}\vec{I}_M$ where $\vec{I}_M$ is the $M \times M$ identity matrix. Hence (\ref{eq:xidef}), (\ref{eq:gmse}), (\ref{eq:gammaestimate}) lead to the characteristic equation of the block system
\begin{equation}
x_i = \frac{L-1}{L}g\stxt{mse}\left(\frac{M-1}{Lx_{i-1}}\right)+\sigma^2\ .
\end{equation}
Using a variable exchange we obtain a form of the characteristic equation which will be central to our analysis, viz.,
\begin{equation}
x'_i = g\stxt{mse}\left(\frac{1}{\alpha' x'_{i-1}}\right)+\sigma'^2
\label{eq:char}
\end{equation}
where
\begin{equation}
x_i' = x_i \frac{L}{L-1}, \quad \quad \sigma'^2 = \sigma^2 \frac{L}{L-1}, \quad \quad \alpha'=\frac{L-1}{M-1} . \label{eq:varexch}
\end{equation}
This representation allows us to carry out a performance analysis of the system for finite values of $L$ and $M$ where the data stream size $N$ is going to infinity for the purpose of allowing any number $i$ of demodulation iterations to be considered as message passing on respective trees (relying on Lemma~\ref{lem:cycle_free}).

Equation (\ref{eq:char}) is initialized with 
\[
x_0' = g\stxt{mse}(0)+\sigma'^2
\]
which constitutes the noise-and-interference power before the demodulation iterations start. Due to signal power normalization (see Section~\ref{subsec:IIIA}) $g\stxt{mse}(0)=1$.

Depending on the value of $\sigma'^2$  (\ref{eq:char}) can have a different number of fixed points (see Section~\ref{sec:roots}) but it always converges to the largest fixed point that is less or equal than the initial variance $1+\sigma'^2$.

 
\subsection{Coupled System}

We consider the regularized delay structure introduced in Section~\ref{subsec:coup} where each data stream is divided into $2W+1$ subsections of length $N\stxt{w}$. Any two interfering data streams share a number of interfering subsections transmitted at the same starting times. 
A superposition of the transmitted data streams forms the received sequence that again consists of subsections of length $N\stxt{w}$. We can, therefore, index the subsections of the received sequence by $t=1,2,3,\cdots$. For the purpose of system analysis we assume that time is given in terms of subsections. For the case of coupled systems the noise-and-interference power and MSE values are not just dependent on the demodulation iteration $i$ but also on the time index $t$ of the respective receive subsection.

In order to asses the noise-and-interference power $x^t_i$ for subsection $t$ at iteration $i$ we look again at equation (\ref{eq:xidef}). All variable nodes connected to a channel node that belongs to receive subsection $t$ are within subsections indexed by $\tau_1 = t-W,t-W+1,\cdots,t+W$. In addition, on average $(L-1)/(2W+1)$ variable nodes $(j',l')$ connected to a single check node belong to each of the $2W+1$ subsections. 
Hence
\begin{align}
x^t_i &\defeq \mathop{\sum_{(j',l') \in \mathcal{J}(t')}}_{\textrm{s.t. } (j',l')\neq (j,l)} \frac{1}{L}\mathbb{E}\left(v_{j',l'} - \hat{v}_{j',l',\tau}^{(i-1)}\right)^2 + \sigma^2  =\frac{L-1}{L(2W+1)}\sum_{\tau_1=t-W}^{t+W} \mu^{\tau_1}_{i-1}+\sigma^2\ .
\label{eq:xidef2}
\end{align}
In turn, to compute the MSE $\mu^{\tau_1}_{i-1}$ for subsection $\tau_1$ at iteration $i-1$ we notice that equation (\ref{eq:muidef}) implies computation of the MSE for symbols of a data stream transmitted at time $\tau-W$ in terms of subsections. The $M-1$ replicas of each data symbol in this data stream experience SNR values $1/(L x^{\tau_2}_{i-1})$, $\tau_2 = \tau_1 - W, \cdots, \tau_2+W$. Again, on average each SNR is experienced by $(M-1)/(2W+1)$ symbols and we obtain
\begin{equation}
\mu^{\tau_1}_{i-1} = g\stxt{mse}\left(\frac{M-1}{2W+1} \sum_{\tau_2=\tau_1-W}^{\tau_1+W} \frac{1}{Lx^{\tau_2}} \right)\ .
\label{eq:mutau1}
\end{equation}
Combining (\ref{eq:xidef2}) and (\ref{eq:mutau1}) together, applying a variable substitution (\ref{eq:varexch}), we obtain the characteristic equation of the coupled system
\begin{equation}
x_i^t  =  \frac{1}{2W+1} \sum_{\tau_1=-W}^{W}  g\stxt{mse} \left(\frac{1}{\alpha'} \frac{1}{2W+1}
         \sum_{\tau_2=-W}^{W}  \frac{1}{x^{t+\tau_1+\tau_2}_{i-1}} \right) + \sigma'^2\ .   
\label{eq:coupsys}
\end{equation}

We assume that the transmission starts at time $t=1$. At every time instant $t$, a new set consisting of $L\stxt{w}$ data streams is transmitted. 
As a result, the initial conditions for recursion (\ref{eq:coupsys}) can be formulated as
\begin{align}
x_0^t &= 0 \quad t \leq 0\ , \label{eq:xt00} \\
x_0^t &= \frac{t}{2W+1}g\stxt{mse}(0) + \sigma'^2 = \frac{t}{2W+1} + \sigma'^2; \quad t \in [1,2W+1] \label{eq:xt001}\\
x_0^t &= g\stxt{mse}(0)  + \sigma'^2 = 1  + \sigma'^2\quad t > 2W+1\ . \label{eq:xt01}
\end{align}
where (\ref{eq:xt001}) indicates the SINR of the data streams equals $0$ before the demodulation starts.

\subsection{Demodulation/Decoding Schedules}

\vspace{3mm}
\noindent{\it Two-Stage Schedule}
\vspace{3mm}

For block systems the residual SINR per symbol after $I$ demodulation iterations equals $\frac{1}{\alpha' x'_{i}}$. Assuming that each data stream is encoded with an identical error correction code with SNR threshold $\theta$ we obtain asymptotically error-free performance if
\[
\frac{1}{\alpha' x'_{i}} \geq \theta\ .
\]
For coupled system the SINR for subsection $t$ after $I$ iterations equals 
\[
\gamma_I^t = \frac{1}{\alpha'(2W+1)}\sum_{j=-W}^{W} \frac{1}{x_I^{t+j}}\ 
\]
from (\ref{eq:coupsys}). Hence the subsequent error correction decoding performed as the second stage of the two-stage receiver is successful if
\begin{equation}
\gamma_I^t \geq \theta\  \quad \textrm{ for } \quad t \geq 1\ .
\label{eq:secstage}
\end{equation}

\vspace{3mm}
\noindent{\it Hard Decoding Feedback and Cancellation Schedule}
\vspace{3mm}

For hard decision decoding and cancellation the receiver performs the decoding of any data stream for which the SINR exceeds the error-correction code threshold $\theta$ and cancels the impact of that stream from the received signal. This means that any data streams with SINR above $\theta$ will no longer contribute to the noise-and-interference power in (\ref{eq:char}) and (\ref{eq:coupsys}). Let us define the MSE function including the error correction decoder as 
\[
 g\stxt{mse,ecc}(\gamma) \defeq
\begin{cases}
    0,& \text{if } \gamma \geq \theta\\
    g\stxt{mse}(\gamma),              & \text{otherwise.}
\end{cases}
\]
As soon as the SINR reaches the error correction code threshold the function sets to $0$ accounting for the fact that the respective data stream is eliminated in the cancellation process. The characteristic equations for the coupled and block systems are modified accordingly and we obtain
\begin{equation}
x'_i = g\stxt{mse,ecc}\left(\frac{1}{\alpha' x'_{i-1}}\right) + \sigma'^2 
\label{eq:charecc}
\end{equation}
and 
\begin{equation}
x_i^t  =  \frac{1}{2W+1} \sum_{\tau_1=-W}^{W}  g\stxt{mse,ecc} \left(\frac{1}{\alpha'} \frac{1}{2W+1}
         \sum_{\tau_2=-W}^{W}  \frac{1}{x^{t+\tau_1+\tau_2}_{i-1}} \right) + \sigma'^2\ .   
\label{eq:coupsysecc}
\end{equation}
The systems converge iff $x'_i$ (or $x_i^t$ for every $t \geq 1$ respectively) converge to $\sigma^2$ after a number of iterations indicating that the residual of the received signal contains just noise while all data streams have been decoded.

\subsection{Fixed Points of the Characteristic Equation}
\label{sec:roots}

 Since the MSE function $g\stxt{mse}(\cdot)$ is differentiable and strictly decreasing~\cite{BurSchShiKry2004,verdu11} the values $x'_i$ in the recursive characteristic equation (\ref{eq:char}) are strictly decreasing with demodulation iterations $i$. Depending of the modulation symbol alphabet $\mathcal{A}$, $\sigma'^2$, and $\alpha'$  (\ref{eq:char}) may have different numbers of fixed points. For each $2^B$-PAM signal alphabet there exists a limiting threshold in terms of the noise power $\sigma^2\stxt{s}(B)$ such that for $\sigma'^2>\sigma^2\stxt{s}(B)$ the characteristic equation (\ref{eq:char}) has a single fixed point $x\stxt{s}$ which depends on the system load $\alpha'$ and the noise power $\sigma'^2$.
 
\begin{figure}[t]
\setlength{\unitlength}{1mm}
\begin{picture}(170,70)
   \put(0,0){\includegraphics{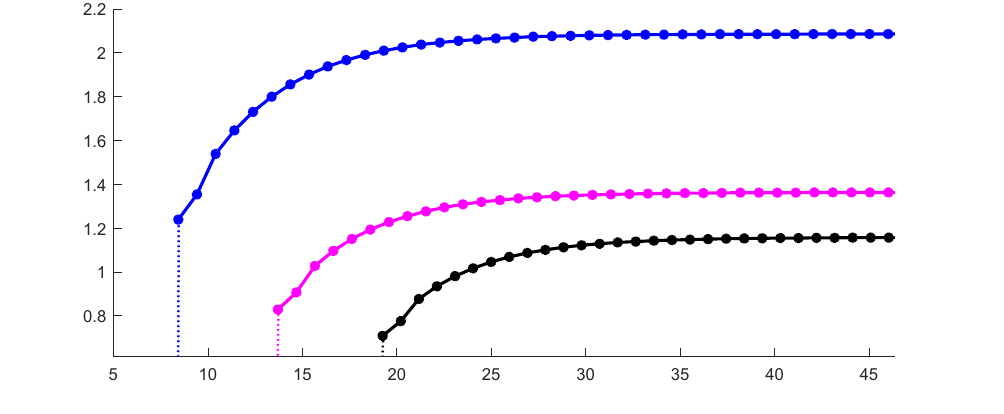}}
   \put(75,2){SNR [dB]}
\end{picture}
\caption{System loads $\alpha\stxt{s}$ for which the characteristic equation (\ref{eq:char}) has exactly two fixed points 2-PAM (blue), 4-PAM (magenta), 8-PAM (black) symbol alphabets are shown as a function of the total system SNR $1/\sigma^2$ in dB. Limiting SNRs $1/\sigma^2\stxt{s}(B)$ for $B=1,2,3$ are given by the blue, magenta, and black dotted vertical lines  respectively. }
\label{Fig:alphas}
\end{figure}

For $\sigma'^2 > \sigma^2\stxt{s}(B)$  (\ref{eq:char}) can have one, two or three fixed points. We can define a system load value $\alpha\stxt{s}(B,\sigma'^2)$ such that for $\alpha' < \alpha\stxt{s}(B,\sigma'^2)$ (\ref{eq:char}) has only one fixed point, two for $\alpha' = \alpha\stxt{s}(B,\sigma'^2)$, and three fixed points for $\alpha' > \alpha\stxt{s}(B,\sigma'^2)$.
Fig.~\ref{Fig:alphas} shows $\alpha\stxt{s}(B,\sigma'^2)$ as a function of the SNR $1/\sigma'^2$ given in dB. The blue curve corresponds to the case $B=1$ (2-PAM), the magenta curve is for $B=2$ (4-PAM), and the black curve is for $B=3$ (8-PAM) symbol alphabets. The threshold SNR levels $1/\sigma^2\stxt{s}(B)$ in decibels are shown by the vertical lines. 


Consider now the 2-PAM case and denote the three fixed points of the system by $x^{(1)}<x^{(2)}<x^{(3)}$. Lemma~\ref{lem:roots} states upper and lower bounds on the fixed points $x^{(1)}$ and $x^{(3)}$. The proposed bounds will be used in the proof of the main result stated in the next section. We note that in case (\ref{eq:char}) has a single fixed point $x\stxt{s}$ it satisfies Lemma~\ref{lem:roots}~(a).

\begin{lemma}
\label{lem:roots}
Consider $\sigma'^2 \leq 1$. Then the following statements are satisfied:\\

\noindent
(a) for $\alpha' \in [0,\mathcal{C}(\sigma'^2)]$
\begin{equation}
\sigma'^2 \leq x^{(1)} \leq (1+e^{-1/\sigma'})\sigma'^2 \leq 2\sigma'^2\ ,
\label{eq:lem1a}
\end{equation}
(b) for $\alpha' \in [4,\mathcal{C}(\sigma'^2)]$
\begin{equation}
 (1+\sigma'^2)^2 - \frac{3}{\alpha'}  \leq \left(x^{(3)}\right)^2 \leq (1 + \sigma'^2)^2\ ,
\label{eq:lem1b}
\end{equation}
(c) for $\alpha' \in [4,\mathcal{C}(\sigma'^2)]$
\begin{equation}
\frac{1}{2}(1 + \sigma'^2) \leq 1+\sigma'^2 - \frac{1}{\alpha'(1+\sigma'^2)}  - \frac{2}{\alpha'^2(1+\sigma'^2)^3} \leq x^{(3)} \leq 1 + \sigma'^2
\label{eq:lem1c}
\end{equation}
 where
\begin{equation}
\mathcal{C}(\sigma'^2) = \frac{1}{2} \log_2 \left(1+\frac{1}{\sigma'^2}\right)\ .
\label{eq:alphahat}
\end{equation}
is the capacity of the AWGN channel with SNR $1/\sigma'^2$. 
\end{lemma}
\begin{proof}
See Appendix~\ref{app:roots}.
\end{proof}

In contrast to the block system (\ref{eq:char}), the fixed points $\vec{x}^*$ of the coupled system (\ref{eq:coupsys}) are vectors. Again we notice that due to the fact that the MSE function is decreasing the components $x_i^t$ of the coupled recursion vectors are decreasing as well for each $t=1,2,3,\cdots$, and  $x^*_t = \lim_{i \rightarrow \infty} x_i^t$, $t=1,2,3,\cdots$. 

We show (see Appendix~\ref{sec:thmproofs}) that the components of the vector fixed points $\vec{x}^*$ are concentrated around the fixed points $x^{(1)}$ and $x^{(3)}$ of the block system characteristic equation (\ref{eq:char}) depending on the parameters $\alpha',\sigma'^2$. The smallest fixed point $x^{(1)}$ is close to the value of the noise power $\sigma^2$ (Lemma~\ref{lem:roots} (a)). In the next section we will determine the rates achievable by block and coupled system and state the main result that the system's performance approaches channel capacity. 


%

\subsection{Achievable Rate and Capacity}
\label{sec:sepeffcap}

For loads $\alpha' > \alpha'\stxt{s}$ the block system's noise-and-interference power converges to the single fixed point of the characteristic equation (\ref{eq:char}) $x\stxt{s}$ and the corresponding SINR to $(\alpha' x\stxt{s})^{-1}$. In order to determine the maximum sum rate achieved by the entire system we assume that error-correction codes used to encode data streams are optimal with respect to the SINR $(\alpha' x\stxt{s})^{-1}$, the symbol constellation alphabet $\mathcal{A}$ and transmission over the AWGN. 
The total communication rate (sum-rate) achievable by the block system in this case equals
\begin{equation}
\mathcal{R}\stxt{u}(\alpha,\sigma^2) = \frac{L}{M} \mathcal{C}_\mathcal{A} \left(\frac{1}{\alpha' x\stxt{s}}\right) = \alpha \mathcal{C}_\mathcal{A} \left(\frac{1}{\alpha' x\stxt{s}}\right) 
\label{eq:achsumrate}
\end{equation}
where $\mathcal{C}_\mathcal{A}(\gamma)$ denotes the capacity of the AWGN channel with input alphabet $\mathcal{A}$ and SNR $\gamma$ and $x\stxt{s}$ is the single fixed point of (\ref{eq:char}). The factor $ \alpha =\frac{L}{M}$ accounts for the number of data streams $L$ in the system and the repetition factor $M$.
We recall that the system (\ref{eq:coupsys}) operates with total signal power $1$ and noise power $\sigma^2$. Therefore, the corresponding capacity of the (real-valued) AWGN channel for these parameters equals
$\mathcal{C}(\sigma^2)$ (where $\sigma^2=\sigma'^2(L-1)/L$). 

For the case when the characteristic equation has three fixed points, i.e., for $\sigma' < \sigma'\stxt{s}$ the block system noise-and-interference power converges to the largest fixed point $x^{(3)}$ (\ref{eq:char})  and the achievable rate of the block system equals
\begin{equation}
\mathcal{R}\stxt{u}(\alpha,\sigma^2) =  \alpha \mathcal{C}_\mathcal{A} \left(\frac{1}{\alpha' x^{(3)}}\right) ,
\label{eq:achsumrate}
\end{equation}

The coupled system, however, converges to the smallest fixed point $x^{(1)}$ of the characteristic equation (\ref{eq:char}) in case the system load $\alpha'$ does not exceed a limiting load $\bar{\alpha}=\bar{\alpha}(\sigma'^2)$ determined via the system's potential function discussed in Appendix~\ref{sec:thmproofs}. For higher loads $\alpha' > \bar{\alpha}$ the coupled system converges to the largest fixed point $x^{(3)}$ of (\ref{eq:char}) just like the block system. Hence the rates achievable by the coupled system are 
\begin{equation}
\mathcal{R}\stxt{coup}(\alpha,\sigma^2) =  \begin{cases}
    \alpha \mathcal{C}_\mathcal{A} \left(\frac{1}{\alpha' x^{(1)}}\right) ,& \text{if } \alpha \leq \bar{\alpha}, W \geq \bar{W} \\
    \alpha \mathcal{C}_\mathcal{A} \left(\frac{1}{\alpha' x^{(3)}}\right) ,               & \text{otherwise}
\end{cases}
\label{eq:achsumrate}
\end{equation}
for sufficiently large window size $\bar{W}$ (see Appendix~\ref{sec:thmproofs}). We are now ready to state the main result of the paper for the case of 2-PAM modulation ($B=1$): 
\begin{theorem}
\label{thm:mainthm}
There exists a $\bar{W}>0$ such that for any $W > \bar{W}$
\begin{equation}
\mathcal{C}(\sigma^2) -\mathcal{R}(\bar{\alpha},\sigma^2) \leq  \mathcal{G}\stxt{asymptotic}(\sigma^2) +  \mathcal{G}\stxt{finite}(L,M,\sigma^2) 
\end{equation}
where
\begin{equation}
 \mathcal{G}\stxt{asymptotic}(\sigma^2)  = \frac{1.081}{\mathcal{C}(\sigma^2)} \quad \quad  \mathcal{G}\stxt{finite}(L,M,\alpha,\sigma^2) = \frac{1}{2}\log_2\left(\frac{\sigma^2+1}{\sigma^2+1-\frac{1}{L}}\right) + \frac{\mathcal{C}(\sigma^2)-1}{M-1}
\end{equation}
for $\mathcal{C}(\sigma^2) \geq 4.26$, and, therefore,
\begin{align}
&\lim_{L,M \rightarrow \infty}  \mathcal{G}\stxt{finite}(L,M,\alpha,\sigma^2) = 0\ , \\  
&\lim_{\sigma^2 \rightarrow 0}  \mathcal{G}\stxt{asymptotic}(\sigma^2) = 0  \ . 
\end{align}
\end{theorem}

\begin{figure}[t]
\setlength{\unitlength}{1mm}
\begin{picture}(160,80)
   \put(0,0){\includegraphics{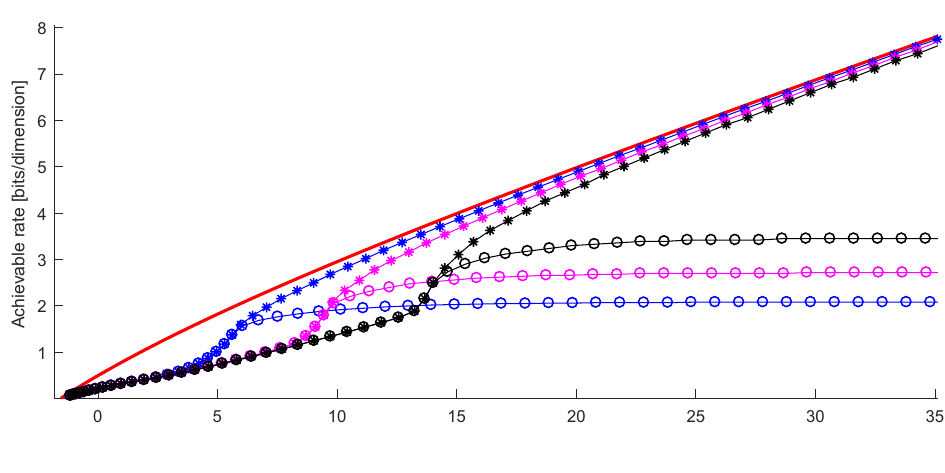}}
   \put(75,2){$E\stxt{b}/N_0$ [dB]}
\end{picture}
\caption{Rates achievable by coupled (stars) and block (circles) systems in comparison to channel capacity (solid red curve). The performances of the systems with 2-PAM, 4-PAM, and 8-PAM symbol alphabets are shown by blue, magenta, and black curves respectively.}
\label{Fig:Rach}
\end{figure}

The proof is given in Appendix~\ref{sec:thmproofs}. The expression in Theorem~\ref{thm:mainthm}\footnote{A slightly tighter bound on the asymptotic gap to capacity can be found in the previous version of this paper posted on Arxiv. It requires a significantly lengthier derivation.} demonstrates the asymptotic behavior of the achievable rate and shows that the gap between the achievable rate and the AWGN channel capacity tends to zero as SNR $1/\sigma^2$ increases. The next theorem is dedicated to the asymptotic gap between the achievable rate and capacity.
\begin{theorem}
\label{thm:scaling}
\begin{equation}
\mathcal{C}(\sigma^2) - \max_\alpha \mathcal{R}_{\textrm coup}(\alpha,\sigma^2) =  \frac{1}{4\ln2 \hspace{1mm} \mathcal{C}(\sigma^2)} + \frac{1}{12\ln2 \hspace{1mm} \mathcal{C}(\sigma^2)^2} + \mathcal{O}\left(\frac{1}{\mathcal{C}(\sigma^2)^3}\right)
\end{equation}
as $\sigma^2 \rightarrow 0$ and $\mathcal{C}(\sigma^2) \rightarrow \infty$.
\end{theorem}

The proof is given in Appendix~\ref{sec:scaling}. The gap, which  is inverse proportional to the capacity itself, is plotted in Fig.~\ref{Fig:GapToCap} (squares) alongside with an upper bound given by Theorem~\ref{thm:mainthm} (circles), asymptotic scaling expression given by Theorem~\ref{thm:scaling} (diamonds), and achievability (lower) bound (stars) discussed below.

Fig.~\ref{Fig:Rach} shows the maximum sum-rates achievable by the block (circles) and coupled (stars) systems for 2-PAM (blue), 4-PAM (magenta), and 8-PAM symbol alphabets as a function of $E\stxt{b}/N_0$. The solid red line shows AWGN channel capacity. We note that in the high SNR regime the rates achievable by the coupled systems approach the capacity curve. Conversely, the rates achievable by the block systems saturate with 8-PAM rate being the highest. This is because the coupled system can converge to the smallest (almost interference-free) fixed point of the characteristic equation while block system always converges to the largest fixed point.
In low-SNR regime both block and coupled system rates coincide as they are determined by the same single fixed point of the characteristic equation.



\begin{figure}[t]
\setlength{\unitlength}{1mm}
\begin{picture}(160,80)
   \put(0,0){\includegraphics{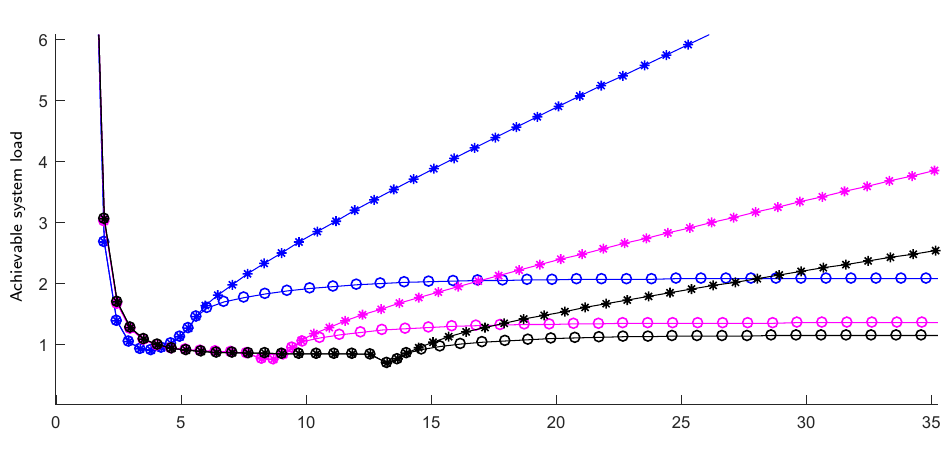}}
   \put(75,2){$E\stxt{b}/N_0$ [dB]}
\end{picture}
\caption{System loads $\alpha$ maximizing the total achievable rate for coupled (stars) and block (circles) systems plotted as functions of $E\stxt{b}/N_0$. Loads for 2-PAM, 4-PAM, 8-PAM symbol alphabets are given by the blue, magenta, and black curves respectively. }
\label{Fig:maxalphas}
\end{figure}

Fig.~\ref{Fig:maxalphas} demonstrates  the system loads maximizing the rates achievable by the block and coupled systems for $2$-PAM, $4$-PAM and $8$-PAM constellation alphabets as a function of $E\stxt{b}/N_0$ (that corresponds to the maximum achievable rate).
For high SNRs the system loads maximizing the rate achievable by the block system coincide with the value $\alpha\stxt{s}(B,\sigma'^2)$, the highest system load such that (\ref{eq:char}) has a single fixed. This is the case since the performance of the block system is driven by the largest fixed point of (\ref{eq:char}). The optimal loads the coupled system coincide with these of block systems for low SNRs. For high SNRs coupled systems can utilize higher loads since coupled systems are driven by the smallest fixed point of (\ref{eq:char}) at high SNRs the coupled system can get advantage of higher system loads.  This behavior is consistent with the behavior of the individually optimal and jointly optimal MAP decoding thresholds studied in~\cite{mulger04,guover05} (using replica method) as discussed in Section~\ref{sec:map}.

\subsection{Bounds on Achievable Rates}
\label{sec:Rbnd}

The proof of the asymptotic capacity-achieving property involves potential functions~\cite{PfisterP14,KRU} which can be used to characterize the conditions for convergence of coupled systems to their fixed points. While the proof in Appendix~\ref{sec:thmproofs} uses potential functions as in~\cite{PfisterP14} here we use a potential function as discussed in~\cite{KRU} to visualize the capacity approaching behavior of the coupled systems for the two demodulation/decoding schedules from Section~\ref{sec:decodingschedules}. We start with the two-stage schedule. We consider the asymptotic scenario in this case where $L,M \rightarrow \infty$; $\alpha=L/M$ is constant. A variable exchange in the characteristic equation (\ref{eq:char}) leads to a characterization of fixed points $u$ that satisfy\footnote{This equation appears in~\cite{KRU} where the system we study is discussed in an example. Here we add further steps relating it to the achievable rate and capacity. }
\begin{equation}
\frac{1}{u} = \alpha g\stxt{mse}(u) + \alpha \sigma^2\ .
\label{eq:charKRU}
\end{equation}

\begin{figure}[t]
\setlength{\unitlength}{1mm}
   \begin{picture}(160,110)
    \put(10,57){\rotatebox{90}{$g\stxt{mse}(\textrm{SNR})$}}
   \put(2,0){\includegraphics{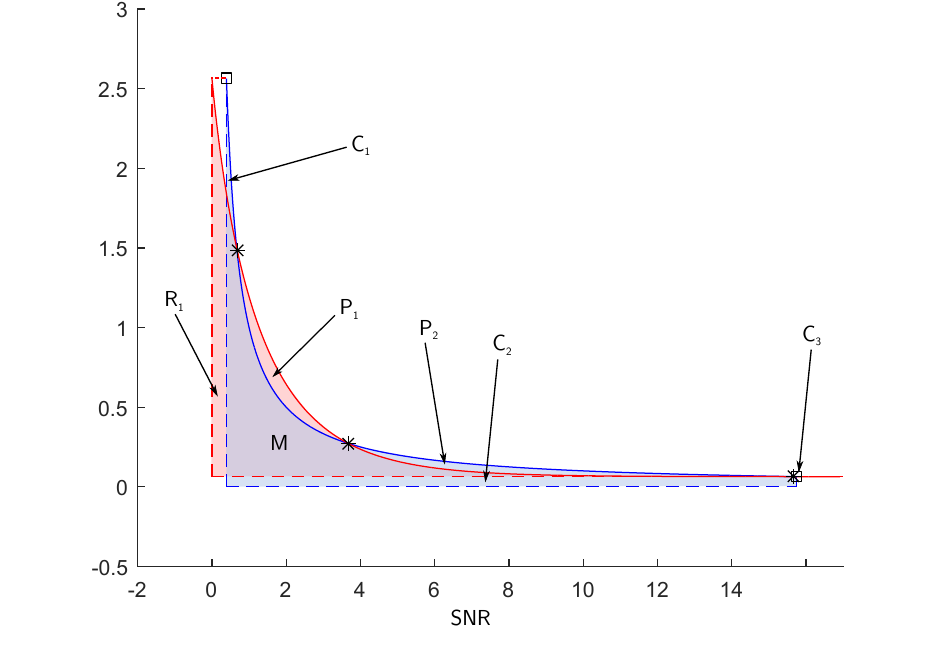}
}
\end{picture}
\caption{Graphic representation of the MSE vs SNR, 2-stage demodulation/decoding.}
\label{Fig:MSESNR}
\end{figure}

Graphically the fixed points of (\ref{eq:charKRU}) are illustrated in Fig.~\ref{Fig:MSESNR}, which shows MSE vs SNR for a 2-PAM system with load $\alpha=2.5$ and $\sigma^2 = 0.0254$ choosen such that the potential function of the coupled system equals $0$. The left hand side of the (\ref{eq:charKRU}) is shown by the blue solid curve while the right hand side of (\ref{eq:charKRU}) is depicted by the red curve. The three black stars depict the three fixed points $(1/(\alpha x^{(3)}),\alpha x^{(3)})$, $(1/(\alpha x^{(2)}),\alpha x^{(2)})$, and $(1/(\alpha x^{(1)}),\alpha x^{(1)})$ of  (\ref{eq:charKRU}). According to~\cite{KRU} the paramater pair $(\alpha, \sigma^2)$ for which the two areas labeled $\vec{P}_1$ and $\vec{P}_2$ between the curves corresponding to the left and right hand sides of the characteristic equation (\ref{eq:charKRU}) are equal 
correspond to the case when the potential function of the system equals $0$. For fixed $\sigma^2$ this corresponds to the highest system load $\alpha$ such that the coupled system (\ref{eq:coupsys}) converges to the smallest (nearly interference-free) fixed point $x^{(1)}$. 

The area colored in light red  (divided by $2\ln 2$) can be computed by integration of the right hand side of (\ref{eq:charKRU}) and equals the total rate achieved by the coupled system
\begin{equation}
\mathcal{R}\stxt{coup}(\alpha,\sigma^2) = \int_0^{1/(\alpha x^{(1)})}  \frac{1}{2\ln 2} \alpha g\stxt{mse}(u) d u\ 
\end{equation}
while the area colored in blue (divided $2\ln 2$) equals the AWGN channel capacity 
\begin{equation}
\mathcal{C}(\sigma^2) =  \frac{1}{2\ln 2}  \int_{1/(\alpha(1+\sigma^2))}^{1/(\alpha \sigma^2)}  \frac{1}{u}   d u\ . 
\label{eq:capmse}
\end{equation}
According to Lemma~\ref{lem:roots} the fixed points are contained between $(1/(\alpha (1+\sigma^2)),\alpha (1+\sigma^2))$ and $(1/(\alpha \sigma^2),\alpha \sigma^2)$. We also notice that the area $\vec{R}_1$ belongs to the achievable rate only. The areas $\vec{C}_1$, $\vec{C}_2$, and $\vec{C}_3$ (a tiny area between $1/(\alpha x^{(1)})$ and $1/(\alpha \sigma^2)$) belong exclusively to the AWGN capacity calculation. Both achievable rate and capacity share the main area $\vec{M}$. A similar situation would happen in case of no-binary symbol alphabets and, as we will see later, in case of other demodulation/decoding schedules where usage of error-correction in the iterative demodulation loop impacts the MSE curve. 

For the pictorial representation we can see that the difference between the achievable rate and capacity can be extracted from the areas  $\vec{R}_1$, 
 $\vec{C}_1$, $\vec{C}_2$, $\vec{C}_3$ where $\vec{R}_1$ and $\vec{C}_2$ are the largest and play the main role. We can derive a simple lower bound on the gap to capacity as
\begin{align}
\mathcal{C}(\sigma^2) &- \mathcal{R}\stxt{coup}(\alpha,\sigma^2) = \frac{1}{2\ln 2}\left(\vec{C}_1+\vec{C}_2+\vec{C}_3-\vec{R}_1\right) \geq \frac{1}{2\ln 2}\left(\vec{C}_2 - \vec{R}_1\right) \nonumber \\
& \geq \frac{1}{2\ln 2}\alpha \sigma^2 \left(\frac{1}{\alpha\sigma^2}-\frac{1}{\alpha(1+\sigma^2)}\right) - \frac{1}{4\ln 2}\frac{1}{\alpha(1+\sigma^2)}\left(\alpha g\stxt{mse}(0)+\sigma^2+\alpha g\stxt{mse}\left(\frac{1}{\alpha(1+\sigma^2)}\right)+\sigma^2\right) \nonumber \\
& =\frac{1}{2\ln 2}\frac{1}{1+\sigma^2} - \frac{1}{2\ln 2}\frac{1}{1+\sigma^2}\left(\frac{1}{2}+\frac{1}{2}g\stxt{mse}\left(\frac{1}{\alpha(1+\sigma^2)}\right) + \frac{\sigma^2}{\alpha}\right) \nonumber \\
&= \frac{1}{2\ln2}\frac{1}{1+\sigma^2}\left(\frac{1}{2}-\frac{1}{2}g\stxt{mse}\left(\frac{1}{\alpha(1+\sigma^2)}\right) - \frac{\sigma^2}{\alpha}\right)\ .
\label{eq:caplb}
\end{align}
For the case of 2-PAM signals we can specialize it further using the upper bound $(1+s)^{-1} \geq g_2(s)$ derived in~\cite{BurSchShiKry2004} and obtain
\begin{multline}
\mathcal{C}(\sigma^2) - \mathcal{R}\stxt{coup}(\alpha,\sigma^2)  \geq \frac{1}{2\ln2(1+\sigma^2)}\left(\frac{1}{2}-\frac{1}{2}\frac{\alpha(1+\sigma^2)}{1+\alpha(1+\sigma^2)} - \frac{\sigma^2}{\alpha}\right) \\ 
= \frac{1}{2\ln2(1+\sigma^2)}\left(\frac{1}{2(1+\alpha(1+\sigma^2))} - \frac{\sigma^2}{\alpha}\right)\ .
\label{eq:lowerb}
\end{multline}
As $\sigma^2 \rightarrow 0$ the right hand side is approximately equal to $\frac{1}{2\alpha} \approx \frac{1}{2\mathcal{C}(\sigma^2)}$ which is consistent with the behavior showcased by Theorem~\ref{thm:mainthm}.
The upper bound on the gap derived in Theorem~\ref{thm:mainthm} involves estimation of all areas $\vec{C}_1$, $\vec{C}_2$, $\vec{C}_3$ and relates estimation of the largest fixed point $x^{(3)}$ of (\ref{eq:char}). 

\subsection{Rate Achievable for Hard Decoding Feedback and Cancellation Schedule}
\label{sec:HD}

Consider now the hard decoding feedback and cancellation schedule. Once during the demodulation process a data stream reaches SNR equal or exceeding the threshold $\theta$ of the error correction code the data is passed to the error correction decoder. The decoder is then capable to perform error-free decoding and the effect of the data stream would be perfectly cancelled once the decoder gives its feedback to the demodulator. 
 
\begin{figure}[t]
\setlength{\unitlength}{1mm}
   \begin{picture}(160,110)
   \put(12,55){\rotatebox{90}{$g\stxt{mse}(\textrm{SNR})$}}
   \put(2,0){\includegraphics{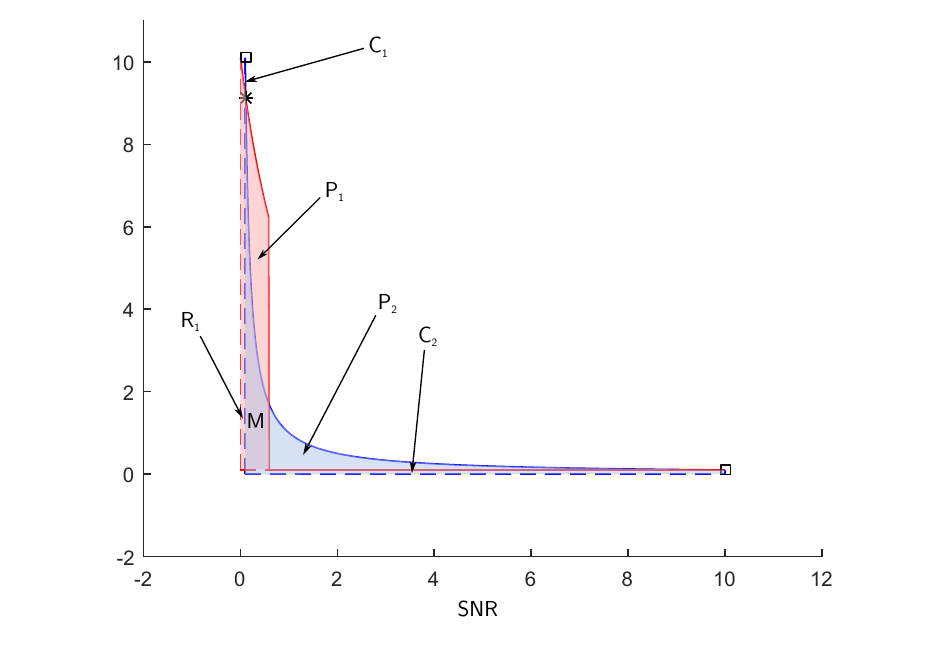}
}
\end{picture}
\caption{Graphical representation of the MSE vs SNR, one time hard decoding feedback and cancellation schedule.}
\label{Fig:MSESNR2}
\end{figure}
 
Fixed points of the modified characteristic equation
 \begin{equation}
\frac{1}{u} = \alpha g\stxt{mse,ecc}(u) + \alpha \sigma^2\ 
\label{eq:charKRUecc}
\end{equation}
as well as the areas between the curves characterizing the achievable rate and capacity are given in Fig.~\ref{Fig:MSESNR2}. The achievable rate expression is now given by
\begin{equation}
\mathcal{R}\stxt{coup,ecc}(\alpha,\sigma^2) = \int_0^{\gamma}  \frac{1}{2\ln 2} \alpha g\stxt{mse}(u) d u 
\end{equation}
while the  AWGN channel capacity is still computed as in (\ref{eq:capmse}). The gap to capacity in terms of the area is now given by
\begin{align}
\mathcal{C}(\sigma^2) - \mathcal{R}\stxt{coup,ecc}(\alpha,\sigma^2) &=   \frac{1}{2\ln 2} \left(\vec{C}_1+\vec{C}_2-\vec{R}_1\right) \geq   \frac{1}{2\ln 2}\left(\vec{C}_2 - \vec{R}_1 \right) \nonumber\\
&= \frac{1}{2\ln2(1+\sigma^2)}\left(\frac{1}{2}-\frac{1}{2}g\stxt{mse}\left(\frac{1}{\alpha(1+\sigma^2)}\right) - \frac{\sigma^2}{\alpha}\right)
\end{align}
which is the same lower bound as in (\ref{eq:caplb}). At the same time we can derive an upper bound 
\begin{align}
\mathcal{C}(\sigma^2) &- \mathcal{R}\stxt{coup,ecc}(\alpha,\sigma^2) =  \frac{1}{2\ln 2} \left( \vec{C}_1+\vec{C}_2-\vec{R}_1 \right)  \\
&\leq \frac{1}{2\ln 2} \frac{1}{2\alpha} \left(\frac{1}{x\stxt{s}}-\frac{1}{1+\sigma^2}\right)\left(\alpha (1+\sigma^2 - x\stxt{s})\right) 
+ \frac{1}{2\ln 2(1+\sigma^2)}\left(\frac{1}{2}-\frac{1}{2}g\stxt{mse}\left(\frac{1}{\alpha(1+\sigma^2)}\right) - \frac{\sigma^2}{\alpha}\right)
\label{eq:eccbnd1}
\end{align}
where $x\stxt{s}$ is the single fixed point of the characteristic equation (\ref{eq:char}). If we focus on the case of binary symbols we can now use Lemma~\ref{lem:roots}~(c) and the lover bound $1-y \leq g_2(y)$ \cite{BurSchShiKry2004} and based on (\ref{eq:eccbnd1}) further obtain
\begin{align}
\mathcal{C}(\sigma^2) - \mathcal{R}\stxt{coup,ecc}(\alpha,\sigma^2) &=  \frac{1}{2\ln 2} \left[ \frac{1}{2}\frac{(1+\sigma^2 -x\stxt{s})^2}{x\stxt{s}(1+\sigma^2)}+ \frac{1}{1+\sigma^2}\left(\frac{1}{2}-\frac{1}{2}\left(1-\frac{1}{\alpha(1+\sigma^2)}\right) -\frac{\sigma^2}{\alpha}\right)\right] \nonumber\\
&\leq \frac{1}{2\ln 2} \left[ \frac{1}{2}\frac{\left(\frac{1}{\alpha(1+\sigma^2)}+\frac{1}{\alpha^2(1+\sigma^2)^3}\right)^2}{\frac{1}{2}(1 + \sigma^2)^2}+ \frac{1-2\sigma^2(1+\sigma^2)}{2\alpha(1+\sigma^2)^2}\right]\nonumber \\
&\leq \frac{1}{2\ln 2} \left[ \frac{3}{\alpha^2(1+\sigma^2)^4}+ \frac{1-2\sigma^2(1+\sigma^2)}{2\alpha(1+\sigma^2)^2}\right] \leq \frac{1}{2\ln 2 \alpha}\ .
\label{eq:eccbnd2}
\end{align}

For the case of the 2-PAM constellation the capacity-achieving property for hard decoding feedback and cancellation schedule has been proven in \cite{TruComLet12}. Here we look at the problem in more detail and prove that in case of a one time feedback and cancellation schedule convergence to channel capacity happens for any fixed $\sigma^2$ when we let $\alpha \rightarrow \infty$. For the two-stage decoding the system is only capacity achieving asymptotically as  $\sigma^2 \rightarrow 0$. We can formulate this result as a theorem.
\begin{theorem}
\label{thm:thm2}
For any $\sigma^2 \leq 0.1$ (necessary for existence of at least one fixed point) and any $\alpha>4$ there exists a $\bar{W}>0$ such that for any $W > \bar{W}$ such that
\begin{equation}
\mathcal{C}(\sigma^2) - \mathcal{R}\stxt{coup,ecc}(\alpha,\sigma^2) \leq  \frac{1}{2\ln 2 \alpha}
\end{equation}
and, therefore, for any $\sigma^2 \leq 0.1$
\[
\lim_{\alpha \rightarrow 0}  \left( \mathcal{C}(\sigma^2)  - \mathcal{R}\stxt{coup,ecc}(\alpha,\sigma^2)\right) = 0\ . 
\]
\end{theorem}

\section{Numerical Results and Relation to MAP Decoding}

\subsection{Numerical Results}
\label{sec:sims}

\begin{figure}[t]
\setlength{\unitlength}{1mm}
\begin{center}
   \begin{picture}(160,70)
   \put(0,0){\includegraphics{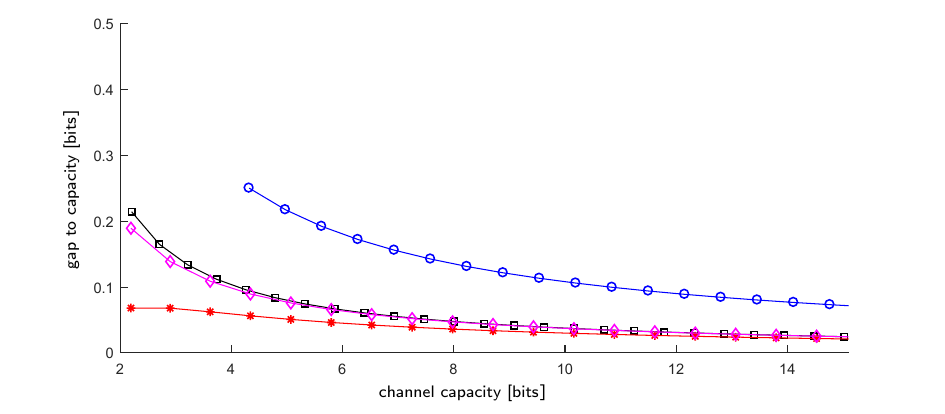}
}
\end{picture}
\end{center}
\caption{The gap between the channel capacity and the total sum-rate achieved by the coupled system with two-stage demodulation/decoding (squares) with 2-PAM symbol constellation plotted as a function of the channel capacity. The upper bound $\mathcal{G}\stxt{asymptotic}(\sigma^2)$ is given by blue circles, the asymptotic scaling expression by magenta diamonds, and the lower bound (\ref{eq:lowerb}) by red stars.}
\label{Fig:GapToCap}
\end{figure}

We start by looking at the gap between the sum-rate achievable by coupled transmission and the AWGN channel capacity. Fig.~\ref{Fig:GapToCap} plots the difference between the capacity of the AWGN channel $\mathcal{C}(\sigma^2)$ and the rate $\mathcal{R}\stxt{coup}(\alpha,\sigma^2)$ achievable by the coupled system with the two-stage schedule in the asymptotic regime $M,L\rightarrow \infty$ (black squares curve). The analytical upper bound $\mathcal{G}\stxt{asymptotic}(\sigma^2)$ (see Theorem~\ref{thm:mainthm}) is given by the blue circles curve. The asymptotic scaling expression of Theorem~\ref{thm:scaling} is given by magenta diamonds curve. Finally, the analytical lower bound (\ref{eq:lowerb}) with $\alpha = \mathcal{C}(\sigma^2)$ is given by the red stars. We notice that all the there curves corresponding to the actual gap-to-capacity and the bounds on it decay as the channel capacity grows. The bounds get tighter as the capacity increases.

\begin{figure}[h]
\setlength{\unitlength}{1mm}
   \begin{picture}(155,90)
   \put(-2,0){\includegraphics{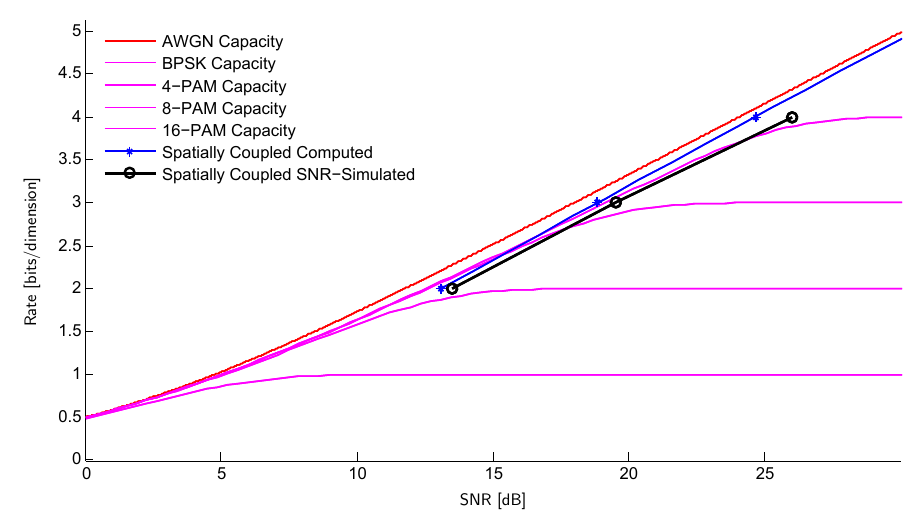}
}
\end{picture}
\caption{Computed (blue) and simulated data rates (black) achievable by coupled data transmission as a function of the total system SNR. The red curve shows the capacity of the AWGN channel with power-constraint input while the magenta curves correspond to the capacities of the AWGN channel with the input constrained to 2,4,8,16-PAM symbol constellations.}
\label{Fig:Sim}
\end{figure}

Simulation results of the spectral efficiency achievable by coupling data transmission are given in Fig.~\ref{Fig:Sim}, which plots the sum-rates achievable by the coupled system with two-stage demodulation/decoding in bits per dimension as a function of the total SNR $1/\sigma^2$ in dB.  
The red curve corresponds to the capacity of the AWGN channel. The magenta curves correspond to capacities of the AWGN channels with constrained inputs where the lower magenta curve is for BPSK (2-PAM) modulation, and the other curves are for 4,8, and 16 PAM constellations (from bottom to top).

The black curve with circles plots simulated sum rate of the coupled data transmission with a block (data stream) length of $MN=10,000$. 
The three simulated points correspond to loads $\alpha=2,3,4$ with parameters $M=250$ (and $L$ chosen accordingly). The coupling window $W=12$ was considered. 
The permutation matrices for each data stream have been chosen randomly with no attempt to eliminate short cycles.  
The SNR at the output of the demodulator is measured and the resulting sum-rate is computed assuming component codes achieve the capacity of the BIAWGN channel for this SNR. We note that the simulated performance curve is close to the theoretically predicted sum-rate performance (blue). Also for each $\alpha=2,3,4$ the simulated performance is closer to the channel capacity than the performance of the corresponding 2,4,8, or 16-PAM capacity curves. Finally, we notice that the performance curve does not saturate at the modulation load of $\alpha =2.07$ as it happens for the block system. The slight divergence at the higher load is arguably caused by the presence of  short cycles in the system's graph which becomes denser as the system load increases. 

\subsection{Relation to MAP Decoding}
\label{sec:map}

Performance limits of joint and separate user decoding in context of multiple access communications have been considered in a number of prior papers. Tse and Verdu showed that in the high-SNR regime joint multi-user detection and decoding of random CDMA signals is capacity achieving~\cite{tsever00}. Tanaka~\cite{tan02} applied the heuristic replica methods from statistical physics to study the spectral efficiencies achieved by iterative multiuser decoding of random CDMA systems in the large system limit. This results were later generalized by Guo and Verdu~\cite{guover05}. M\"uller and Gerstacker quantified the difference between the optimal rates achievable by joint and separate decoding of CDMA signals~\cite{mulger04}.

Before we compare the results achievable by our system under consideration with these achievable with optimal MAP detection and decoding of multi-user signals we would like to mention the two important differences in the system design. Our system model discussed in Section~\ref{sec:sys} includes repetition and permutation of signals at the modulation stage and then it includes coupling of signals at the receiver. The other important feature of our approach is the signal reception in which iterative detection is followed by individual ECC decoding. Our main contribution is a technique to approach the channel capacity using a low-complexity iterative detection followed by individual error correction decoding,  without prohibitively complex joint detection/decoding or MAP demodulation of multiuser signals. 
In the modulation format we consider each data bit is replicated $M$ times and these replicas are spread among the entire block. This is different from random CDMA models considered prior works and is  helpful in case of belief propagation decoding on the modulation graphs. Different bit replicas with different SNRs when combined for data estimation deliver a performance that approaches the performance of MAP demodulation due to the threshold saturation effect obtained via spatial graph coupling.
 
Often used for characterization of a multi-user detection performance the multi-user efficiency parameter defined in~\cite{ver86} is a ratio of the input and output SNR $\eta = \frac{\sigma^2}{\sigma_\infty^2}$. For optimum joint decoding (of CDMA signals)~\cite{tan02} $\eta$ satisfies 
\begin{equation}
\frac{1}{\eta} = 1+ \alpha \textrm{ snr } g_2\left(\eta \textrm{ snr}\right) = 1+ \frac{1}{\sigma^2} g_2\left(\frac{\eta}{\alpha \sigma^2}\right)
\label{eq:chartan}
\end{equation}
where $\textrm{snr} =  1/(\alpha \sigma^2)$ (see also~\cite{guover05} equation (34)). In case of multiple fixed points of the above equation it is  shown in~\cite{guover05}  that $\eta$ resulting the smallest sum-rate
\begin{equation}
C\stxt{joint} (\eta) = \frac{\eta-1- \log \eta}{\sigma^2} + \alpha \mathcal{C}\stxt{BIAWGN}\left(\frac{\eta}{\alpha \sigma^2}\right)
\label{eq:cj}
\end{equation}
is the correct fixed point to choose. The equation (\ref{eq:cj}) also appears in \cite{mulger04}, \cite{guover05} as a characterization of joint multiuser detection. The above characterization  is, however,  a conjecture derived via the informal replica method. The capacity of separate decoding and detection assuming optimum MAP multiuser decoding, again using the replica method analysis, is given by
\begin{equation}
C\stxt{sep} (\eta) =  \alpha \mathcal{C}\stxt{BIAWGN}\left(\eta \textrm{ snr}\right) = \alpha \mathcal{C}\stxt{BIAWGN}\left(\frac{\eta}{\alpha \sigma^2}\right)
\label{eq:csep}
\end{equation}
as derived in~\cite{mulger04}. The gap between joint and separate detection is also given by
\begin{equation}
\frac{\eta-1- \log \eta}{\sigma^2}\ .
\label{eq:gapjs}
\end{equation}
The above equations have been generalized in~\cite{guover05} for the case of unequal user power distributions.

We notice that equation (\ref{eq:chartan}) coincides with  (\ref{eq:char}), the characteristic equation for the block system. 
In this paper we have proven that the coupled system can achieve the smallest fixed point $x^{(1)}$ of  (\ref{eq:char}) with iterative detection for loads $\alpha$ close to the capacity of the MAC channel. The resulting achievable spectral efficiency of our coupled system (\ref{eq:achsumrate}) formula coincides with (\ref{eq:csep}). We also prove that our system design together with spatial coupling leads to an achievable $\eta =\sigma^2/x^{(1)}$ that rapidly approaches $1$ as variance $\sigma^2$ of the noise diminishes (Lemma 2 a)). In such operational regime with $\eta \approx 1$ the gap between the performance of joint and separate decoding schedules (\ref{eq:gapjs}) disappears and the capacity can be approached with a two-stage schedule. 
 

\section{Conclusion}
\label{sec:conc}

We consider modulation of information in the form of a superposition of independent equal-power and equal-rate data streams. Each stream is formed by repetition and permutation of data and the streams are summed up with an offset initiating the effect of ``stream coupling''. The convergence of block and coupled version of the system is studied for various symbol constellation alphabets. We have proven that the proposed system used with iterative demodulation followed by external error control decoding  achieves the capacity of the AWGN channel and the Gaussian multiple-access channel asymptotically and computed the gap to capacity as a function of the system's parameters. In addition, we show that for the hard feedback and cancellation schedule the gap to capacity is smaller than for two-stage schedule and can vanish at finite SNR.



\appendix

\subsection{Proof of Lemma~\ref{lem:cycle_free}}
\label{app:cycle_free}

\begin{figure}[h!]
\setlength{\unitlength}{1mm}
\begin{center}
   \begin{picture}(165,50)
   \put(0,0){\includegraphics{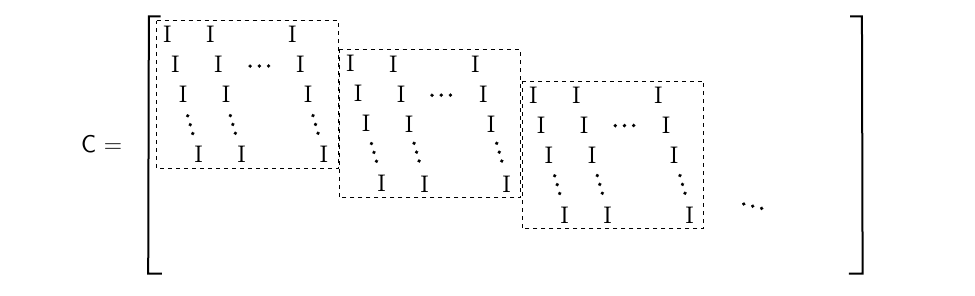}
}
\end{picture}
\end{center}
\caption{Coupling Matrix.}
\label{Fig:Matr}
\end{figure}

We start by observing the fact that each modulated data stream $\tilde{\vec{v}}_l$ can be represented as a product of the encoded data sequence $\vec{v}_l$ and a binary matrix $\vec{H}_l$ which represents repetition and permutation operations,
\[
\tilde{\vec{v}}_l = \vec{H}_l \vec{v}_l\ ,
\]
i.e., $\vec{H}_l=\vec{P}_l \vec{R}$, where $\vec{P}_l$ is an $MN \times MN$ binary permutation matrix specific to data stream $l$, and $MN \times N$ matrix $\vec{R}$ is a binary repetition matrix, repeating each bit of $\vec{v}_l$ $M$ times,
\begin{equation}
\vec{R} = 
\left( \begin{array}{cccccccccccccccc}
1 & 1 & 1 & \cdots & 1 &  &  &  & &        &    &    &  &        &  &          \\
  &    &    &          &   & 1 & 1 & 1& \cdots & 1  &  &    &    &   &  &      \\
  & & & & & & & & & &\ddots  & & & & & \\
  &    &    &       &    &    &   &  &      &   & & 1 & 1 & 1& \cdots & 1  
\end{array} \right)^\textrm{T}
\label{eq:R}
\end{equation}
Without loss of generality we consider $M_1=M$, $M_2=1$ and the regularized coupling structure of the data streams described in Section~\ref{subsec:coup}. The resulting coupled modulated sequence equals 
\[
\vec{s} = \vec{H}\vec{v} =  \vec{C}\vec{P}\tilde{\vec{R}} \vec{v},
\]
where $\vec{v} = (\vec{v}_1,\vec{v}_2,\cdots)$ is the composite  vector of all data streams, and $\vec{C}$ is the coupling matrix. The coupling matrix $\vec{C}$ shown in Fig.~\ref{Fig:Matr} consists of identical sub-matrix blocks depicted by the dashed rectangles. Each block is filled with $(2W+1) \times (2W+1)$ identity matrices $\vec{I}$ of size $N\stxt{w} \times N\stxt{w}$ as shown in  Fig.~\ref{Fig:Matr}. Each consequent block is shifted down by the size of one identity matrix.
The permutation matrix $\vec{P}$ is a block-diagonal matrix which consists of permutation matrices of individual data streams $\vec{P} = \textrm{diag} \left(\vec{P}_1,\vec{P}_2,\vec{P}_3,\vec{P}_4, \cdots\right) $ and the repetition matrix $\tilde{\vec{R}}$ is constructed as given in (\ref{eq:R}) but is infinitely long. 
Here we leave the signature sequence out of the consideration since they do not affect the structure of the system's graph.

\begin{figure}[h]
\setlength{\unitlength}{1mm}
   \begin{picture}(155,90)
   \put(-2,0){\includegraphics{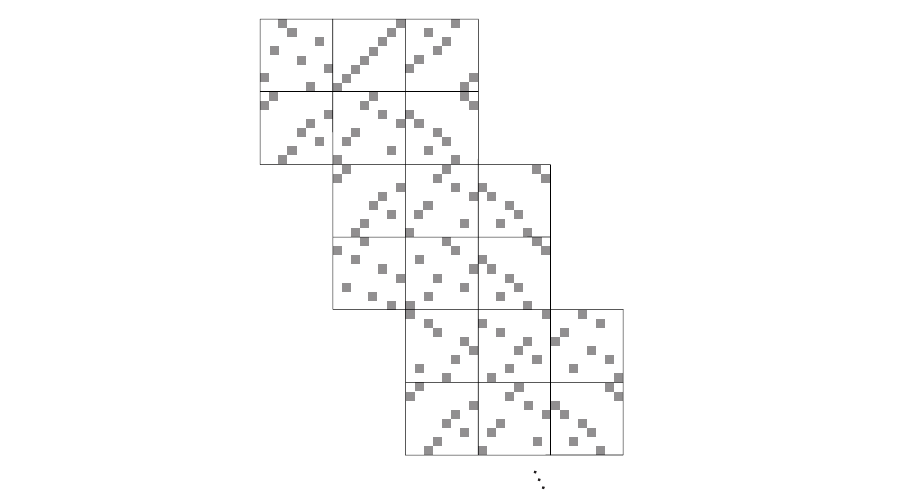}
}
\end{picture}
\caption{Multiple convolutional permutor matrix.}
\label{Fig:ConvPerm}
\end{figure}

The resulting matrix $\vec{H}$ is a band-diagonal binary matrix with $M$ ones in each row and $L$ ones in each column. Such matrices belong to a class of syndrome formers (transposed infinite parity-check matrices)  of LDPC convolutional codes~\cite{ldcc_2}. An example of a syndrome former of a $(3,6)$-regular convolutional LDPC code constructed using $\vec{P}$ that consists of smaller permutation matrices is given in Fig.~\ref{Fig:ConvPerm}.
 
Hence, it is interesting to note that if the addition of the binary modulated data streams was performed as a modulo $2$ addition the resulting coupled sequence would be a codeword of an $(L,M)$-regular convolutional (spatially coupled) LDPC code. Such code can be graphically represented by an infinitely-long  bipartite Tanner graph~\cite{fz99}\cite{ldcc_2}. The same bipartite graph describes modulated data in our system (see Fig.~\ref{Fig:VarGraph} b)). For the case of block system several researchers have shown that it is possible to construct bipartite graphs of block LDPC codes without any cycles of length $2I$ for any fixed integer $I$~\cite{Gal63}(Appendix B), \cite{mar82} if the block length of the code is chosen large enough. A counterpart of this result for convolutional LDPC codes has been shown in~\cite{isit2001}\cite{ldcc_2} where, it has been proven that regular LDPC convolutional codes with graphs of any given girth $I$ can be constructed from block LDPC code graphs of large girths by an unwrapping procedure~\cite{isit2001}\cite{ldcc_2}. This directly implies that modulation graphs of any given girth $I$ can be constructed if $MN$ is chosen to be sufficiently large.


We note that the operation of symbol estimation (\ref{eq:vestimate}) (see Fig.~\ref{Fig:VarGraph}) and the operation of interference cancellation (\ref{eq:ici}) both can be regarded as message message-passing operations on the system's graph. If the system's graph has no cycles of length smaller or equal to $2I$ demodulation with $I$ iterations performed to estimate each bit $v_{j,l}$ is operating on a tree. This proves the lemma. 
We note here that construction of modulation graphs with large girth is done for the purpose of analysis while in practice it is typically enough to expurgate cycles of length four and six.

\subsection{Proof of Lemma~\ref{lem:roots}}
\label{app:roots}
Let us define 
\begin{equation}
h(x) \defeq g_2\left(\frac{1}{\alpha' x} \right) + \sigma'^2\ \geq \sigma'^2\ ,
\label{eq:startsig}
\end{equation} 
where the inequality follows from $g_2(y)>0$, for $y\in (0,\infty)$. Similarly the fact that $g_2(\cdot)$ is non-negative implies the lower bound in Lemma~\ref{lem:roots}~(a) 
\begin{equation}
\sigma'^2 \leq g_2\left(\frac{1}{\alpha' x^{(1)}} \right) + \sigma'^2 =  x^{(1)}\ .
\end{equation}
We use the upper bound on  $g_2(\cdot)$ derived in~\cite{BurSchShiKry2004}
\begin{equation}
g_2(y) \leq \pi Q(\sqrt{y}) = \pi \int_{\sqrt{y}}^{\infty} \frac{1}{\sqrt{2 \pi}} e^{-\frac{z^2}{2}} d z
\label{eq:g2_ub}
\end{equation}
and the condition $\alpha' \leq \mathcal{C}(\sigma'^2)$ of Lemma~\ref{lem:roots} (a) to upper bound
\begin{align}
h(2\sigma'^2)  & \leq \pi Q\left(\frac{1}{\sqrt{\mathcal{C}(\sigma'^2) 2\sigma'^2}}\right)+\sigma'^2 \leq \pi \frac{1}{\sqrt{2 \pi }} \sqrt{\mathcal{C}(\sigma'^2) 2\sigma'^2} \exp \left(-\frac{1}{\mathcal{C}(\sigma'^2) \sigma^2}\right) + \sigma'^2 \label{eq:end2sig}\\
&=\exp  \left(-\frac{1}{\mathcal{C}(\sigma'^2) \sigma'^2} + \frac{1}{2} \ln \frac{\pi}{2} + \frac{1}{2}\ln (2\sigma'^2 \mathcal{C}(\sigma'^2)) \right) + \sigma'^2\\
&= \exp  \left(-\frac{1}{\sigma'^2\frac{1}{2\ln 2}\ln(1+1/\sigma'^2) } + \frac{1}{2} \ln \frac{\pi}{2} + \frac{1}{2}\ln \left(2\sigma'^2 \frac{1}{2\ln 2}\ln(1+1/\sigma'^2)\right) \right) + \sigma'^2 \label{eq:s2}\\
&\leq \exp  \left(-\frac{1}{\sigma'} + 2 \ln(\sigma') \right) + \sigma'^2  < 2\sigma'^2\ . \label{eq:s3}
\end{align}
The first inequality in (\ref{eq:s3}) is valid for $\sigma'^2 \leq 1$ and is due to the fact that the term $-1/\sigma'^2$ dominates the exponent in (\ref{eq:s2}) as $\sigma'^2 \rightarrow 0$. Bounds (\ref{eq:startsig}) and (\ref{eq:end2sig})--(\ref{eq:s3}) imply the existence of a fixed point $\sigma'^2 \leq x^{(1)} \leq 2\sigma'^2$ for any $\alpha' \in [0,\mathcal{C}(\sigma'^2)]$ when $\sigma'^2 \leq 1$. The fixed point $ x^{(1)}$ then satisfies the first inequality in (\ref{eq:s3}) which is the upper bound given by (\ref{eq:lem1a}). Lemma~\ref{lem:roots}~(a) is proved. \qed

The upper bound on $x^{(3)}$ in (\ref{eq:lem1b}) is straightforward since $g_2(y)\leq 1$ for $y \in [0,\infty]$ and, therefore,
\[
x^{(3)} = g_2\left(\frac{1}{\alpha' x^{(3)}} \right) + \sigma'^2 \leq 1 + \sigma'^2\ .
\]
To prove the lower bound  in (\ref{eq:lem1b}) we use the lower bound derived in~\cite{BurSchShiKry2004}
\begin{equation}
g_2(y)\geq 1-y, \quad \quad  y \in [0,\infty]\ .
\label{eq:1ybnd}
\end{equation}
Inequality (\ref{eq:1ybnd}) implies that $x\stxt{r}$, the largest fixed point of the equation
\begin{equation}
x = 1 - \frac{1}{\alpha' x} + \sigma'^2\ ,
\label{eq:xrequat}
\end{equation}
is a lower bound on $x^{(3)}$. Hence
\begin{align}
\left(x^{(3)}\right)^2 \geq \left(x\stxt{r}\right)^2 &= \left[\frac{1+\sigma'^2}{2}\left(1+\sqrt{1 - \frac{4}{\alpha'(1+\sigma'^2)^2}}\right) \right]^2
\\
 &\geq  \frac{\left(1+\sigma'^2\right)^2}{4} \left(2 - \frac{4}{\alpha'(1+\sigma'^2)^2}+2\sqrt{1 - \frac{4}{\alpha'(1+\sigma'^2)^2}}\right) \nonumber \\
 & \geq \frac{\left(1+\sigma'^2\right)^2}{2} - \frac{1}{\alpha'}+\frac{\left(1+\sigma'^2\right)^2}{2}\left(1 - \frac{4}{\alpha'(1+\sigma'^2)^2}\right) =\left(1+\sigma'^2\right)^2-\frac{3}{\alpha'}\ . \label{eq:rootx3i}
\end{align}
Since the value inside the square root  in (\ref{eq:rootx3i}) needs to be non-negative the bound can be used for
\[
\frac{4}{(1+\sigma'^2)^2} \leq \alpha'\ ,
\]
i.e.  for $\alpha' \geq  4$ in particular. Lemma~\ref{lem:roots}~(b) is proved.

To prove the lower bound  in (\ref{eq:lem1c}) (Lemma~\ref{lem:roots}~(c)) we simply refine the bound (\ref{eq:rootx3i}) on $x\stxt{r}$ via a Taylor series expansion of $\sqrt{1 - \frac{4}{\alpha'(1+\sigma'^2)^2}}$. Lemma~\ref{lem:roots} is proved. \qed

\subsection{Proof of Theorem~\ref{thm:mainthm}}
\label{sec:thmproofs}


\begin{proof}

We start with deriving a convergence condition for the coupled system (\ref{eq:coupsys}) using the method of potential functions~\cite{PfisterP14}. 
The use of potential functions for coupled systems was suggested in~\cite{TakeuchiTK12,ttk12perf}, where the authors showed that these functions are related to the Bethe free energy of continuous dynamical system describing a coupled model of a code-division multiple-access (CDMA) system and in \cite{yjnp12,KRU}, where the technique was designed for more general types of coupled recursions. The potential function has also been used in~\cite{dmj11}  to study a  coupled dynamical system describing reconstruction for compressed sensing.

The main part of the proof is dedicated to the derivation of the limiting load $\alpha^*$  for which the coupled system converges to the nearly interference-free state. To do so we use a relation between the mutual information and the MSE~\cite{gsv05} to derive an analytically tractable lower bound on the minimum of the potential function. Finally, we use bounds on the fixed points $x^{(1)}$ and $x^{(3)}$ of the convergence equation, provided by Lemma~\ref{lem:roots}, to compute the achievable communication rate $\mathcal{R}(\alpha^*,\sigma^2)$, prove that it is within a small gap from the channel capacity $\mathcal{C}(\sigma^2)$, and derive the asymptotic behavior of this gap.

We focus on the case of 2-PAM symbol constellation and for a fixed pair of parameters $\alpha',\sigma'^2$ such that $\alpha' \in [4,\mathcal{C}(\sigma'^2)]$  consider the recursion
\begin{equation}
x_{i} = g_2\left(\frac{1}{\alpha' (x_{i-1}+x^{(1)})}\right) + \sigma'^2 - x^{(1)}
\label{eq:eqsys}
\end{equation}
equivalent to (\ref{eq:char}) in a sense that  the fixed points of (\ref{eq:eqsys}) are equal to these of (\ref{eq:char}) reduced by $x^{(1)}=x^{(1)}(\alpha',\sigma'^2)$ which is the smallest fixed point of (\ref{eq:char}). The fixed points of (\ref{eq:eqsys}) are given by  $0,x^{(2)}-x^{(1)}$, and $x^{(3)}-x^{(1)}$.
The coupled recursion corresponding to (\ref{eq:eqsys}) can be written as
\begin{equation}
x_i^t  =  \frac{1}{2W+1} \sum_{\tau_1=-W}^{W}  g_2 \left(\frac{1}{\alpha'} \frac{1}{2W+1}
         \sum_{\tau_2=-W}^{W}  \frac{1}{x^{t+\tau_1+\tau_2}_{i-1}+x^{(1)}} \right) + \sigma'^2 - x^{(1)}\ ,    \quad t > 0, i > 0
\label{eq:coupsysshift}
\end{equation}
and is, in turn, equivalent to (\ref{eq:coupsys}) from the convergence perspective. The resulting vector fixed points of (\ref{eq:coupsys}) and (\ref{eq:coupsysshift}) also differ by $x^{(1)}$. The recursion (\ref{eq:eqsys}) can be stated in the form 
\begin{equation}
x_i = f(g(x_{i-1}),\alpha')
\label{eq:potfunrec0}
\end{equation}
where 
\begin{align}
f(x,\alpha) &\defeq g_2\left(\frac{1}{\alpha'} \left(\frac{1}{x^{(1)}} - x\right)\right) + \sigma'^2 - x^{(1)}\ , \label{eq:fxalp} \\
g(x) &\defeq \frac{1}{x^{(1)}} - \frac{1}{x^{(1)}+x}\ \label{eq:gx}.
\end{align}
The potential function of the block system~(\ref{eq:potfunrec0}) is given~\cite{PfisterP14} by
\begin{align}
U(x,\alpha') &= \ln \frac{x+x^{(1)}}{x^{(1)}}- \frac{\sigma'^2 x}{x^{(1)}(x+x^{(1)})} - \alpha' \int^{\frac{1}{\alpha' x^{(1)}}}_{\frac{1}{\alpha'(x+x^{(1)})}} g_2 (y) d y\ .
\label{eq:uxalpdef}
\end{align}
and based on the fact that $f,g$ are increasing and differentiable (admissibility conditions) we can utilize Theorem 1~\cite{PfisterP14} which states that the fixed points of (\ref{eq:eqsys}) satisfy
\[
\max_{t \geq 0} x^*_t \leq \max \left( \arg \min_x U(x,\alpha') \right)\ .
\]
for sufficiently large $W$. Hence, if for $\alpha',\sigma'^2$ the potential function $U(x,\sigma^2)=0$ for $x=0$ and $U(x,\sigma'^2)>0$ for $x>0$ the components of the fixed point vector $x^*$ of the original coupled system (\ref{eq:coupsys}) do not exceed $x^{(1)}$. 
Let us define 
\begin{equation}
\alpha^*(\sigma'^2) = \sup \left\{ \alpha : \min_{x \in [0,1+\sigma^2]} U(x,\alpha') \geq 0 \right\}\ .
\label{eq:alpstardef}
\end{equation}
For any $\alpha' < \alpha^*$ (\ref{eq:coupsysshift}) converges to $\vec{0}$ and, therefore,  the original coupled system (\ref{eq:coupsys}) converges to a fixed point vector with components not exceeding $x^{(1)}$. 
We will focus of estimating $\alpha^*$ and the respective load $\mathcal{R}\stxt{coup}(\alpha^*,\sigma'^2)$ which we now know is achievable.

The function $U(x,\alpha')$ is plotted in Fig.~\ref{Fig:Uplot} for $\sigma'^{2} = 0.0129$, $0.003347$, and $0.000855$ and the corresponding $\alpha^* =3,4,5$.  Note that  $\alpha^*$ is a decreasing function of $\sigma'^2$.

\begin{figure}[h]
\setlength{\unitlength}{1mm}
   \begin{picture}(155,100)
   \put(0,0){\includegraphics{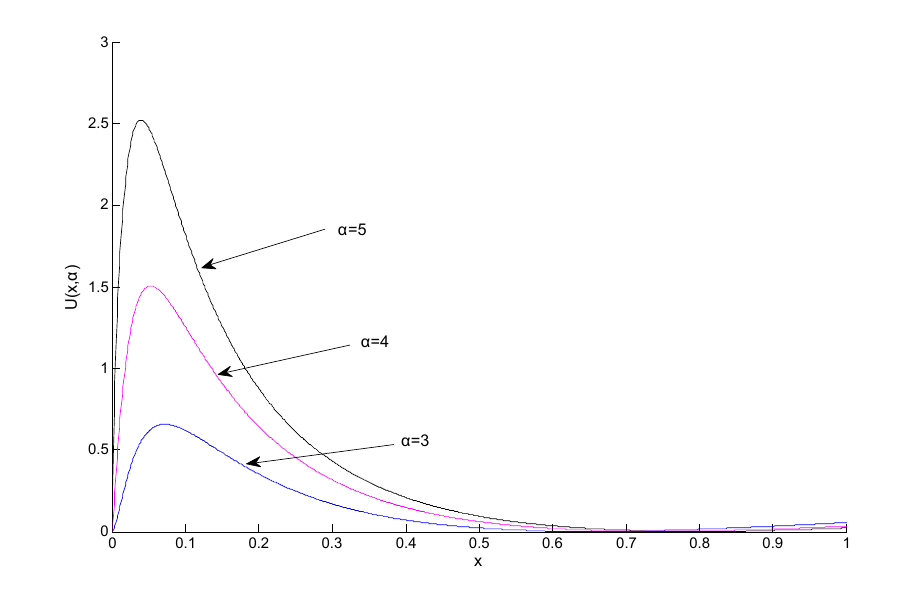}
}
\end{picture}
\caption{Plot of $U(x,\alpha')$ for $\alpha'=3$ and $\sigma'^{2} = 0.0129$ (blue curve),  $\alpha'=4$ and $\sigma'^{2} = 0.003347$ (magenta curve), and $\alpha'=5$ and $\sigma^{2} = 0.000855$ (black curve).}
\label{Fig:Uplot}
\end{figure}

To find the minimum of  $U(x,\alpha')$ defined by (\ref{eq:uxalpdef}) for a given $\alpha'$ and $\sigma'^2$ we compute its partial derivative with respect to $x$ and equate it to $0$
\begin{equation}
\frac{\partial U(x,\alpha')}{\partial x} = \frac{1}{x+x^{(1)}} - \frac{\sigma'^2}{(x+x^{(1)})^2} - g_2\left(\frac{1}{\alpha'(x+x^{(1)})}\right) \frac{1}{(x+x^{(1)})^2} = 0\ .
\end{equation}
which is equivalent to
\begin{align}
x+x^{(1)} &= g_2\left(\frac{1}{\alpha'(x+x^{(1)})}\right) + \sigma'^2\ .
\label{eq:udereq}
\end{align}
Since we consider $\alpha' \in (4,\mathcal{C}(\sigma'^2))$ and $4>\alpha'\stxt{s}$ the equation (\ref{eq:udereq}) has three fixed points. This implies that  $U(x,\alpha')$ has two local minima achieved at $x=0$ and  $x\stxt{m}=x^{(3)}-x^{(1)}$.  Moreover, for $\alpha' =\alpha^*$ we have $U(x_m,\alpha^*)=0$. 
Computing $U(x\stxt{m},\alpha')$ leads to
\begin{align}
U(x\stxt{m},\alpha') =U(x^{(3)}-x^{(1)},\alpha') &= \ln \frac{x^{(3)}}{x^{(1)}} -  \frac{x^{(3)}-x^{(1)}}{x^{(1)}x^{(3)}} \sigma'^2 -  \alpha \int \limits^{\frac{1}{\alpha' x^{(1)}}}_{\frac{1}{\alpha' x^{(3)}}} g_2 (y) d y \defeq u(\alpha',\sigma'^2)\ .
\label{eq:Ualph}
\end{align}

We will now focus on the function $u(\alpha',\sigma'^2)$, compute lower bounds for it and fund out when these are positive in order to find a lower bound on $\alpha^*(\sigma'^2)$ and the achievable rate $\mathcal{R}(\alpha,\sigma^2)$.

The relationship between MSE (i.e. function $g_2(\cdot)$) and the mutual information for BIAWGN channel, derived in~\cite{gsv05} implies  
\begin{equation}
\int \limits^{\frac{1}{\alpha' x^{(1)}}}_{\frac{1}{\alpha' x^{(3)}}} g_2 (y) d y = 2\ln 2 \left( \mathcal{C}\stxt{BIAWGN} \left(\frac{1}{\alpha' x^{(1)}}\right) - \mathcal{C}\stxt{BIAWGN} \left(\frac{1}{\alpha' x^{(3)}}\right)\right)\ .
\label{eq:mmseI}
\end{equation}
Applying (\ref{eq:mmseI}) to (\ref{eq:Ualph}) leads to 
\begin{align}
u(\alpha',\sigma'^2) &= \ln \frac{x^{(3)}}{x^{(1)}} -  \frac{1}{x^{(1)}} \sigma'^2 +  \frac{1}{x^{(3)}} \sigma'^2 -2\ln2 \alpha'  \mathcal{C}\stxt{BIAWGN} \left(\frac{1}{\alpha' x^{(1)}}\right) +2\ln2\alpha' \mathcal{C}\stxt{BIAWGN} \left(\frac{1}{\alpha' x^{(3)}}\right)\ 
\label{eq:Ualph2}
\end{align}
For a given SNR (or equivalently $\sigma'^2$) in order to find a suitable $\alpha'$ such that $u(\alpha',\sigma'^2)>0$ we first find a convenient lower bound on (\ref{eq:Ualph2}). We use the bounds on the capacity of the BIAWGN channel~\cite{Gal06}\cite{Alen96}
\begin{equation}
\frac{\gamma-\gamma^2}{2\ln 2} \leq \mathcal{C}\stxt{BIAWGN} \left(\gamma\right) \leq 1\ .
\label{eq:gamma}
\end{equation}
where the lower bound in (\ref{eq:gamma}) is valid for $\gamma < 1$.  Lemma~\ref{lem:roots} (b) implies that the condition $1/(\alpha' x^{(3)})<1$ is satisfied for $\alpha'>4$.  Applying (\ref{eq:gamma}) to (\ref{eq:Ualph2}) leads to
\begin{align}
u(\alpha',\sigma'^2) &\geq \ln \frac{x^{(3)}}{x^{(1)}} -  \frac{1}{x^{(1)}} \sigma'^2 +  \frac{1}{x^{(3)}} \sigma'^2 -2\ln2 \alpha' +\alpha' \left[\frac{1}{\alpha' x^{(3)}}- \frac{1}{(\alpha' x^{(3)})^2}\right] 
\label{eq:Ualph3}
\end{align}
We then extract the capacity term 
\begin{align}
u(\alpha',\sigma'^2) &\geq  2\ln2\left[\frac{1}{2}\log_2\frac{1+\sigma'^2}{\sigma'^2}\right] +\ln\frac{x^{(3)}}{1+\sigma'^2} + \ln \frac{\sigma'^2}{x^{(1)}}\\ 
& \quad \quad -  \frac{1}{x^{(1)}} \sigma'^2 +  \frac{1}{x^{(3)}} \sigma'^2 -2\ln2 \alpha' +\alpha' \left[\frac{1}{\alpha' x^{(3)}}- \frac{1}{(\alpha' x^{(3)})^2}\right] \\
&\geq  2\ln2\left[\mathcal{C}(\sigma'^2)-\alpha'\right] +\ln\frac{x^{(3)}}{1+\sigma'^2} + \ln \frac{\sigma'^2}{x^{(1)}}
 -  \frac{1}{x^{(1)}} \sigma'^2 +  \frac{1}{x^{(3)}} \sigma'^2  + \frac{1}{x^{(3)}}- \frac{1}{\alpha'( x^{(3)})^2}\
\label{eq:Ualph4}
\end{align}
At the next step we use Lemma~\ref{lem:roots} to  bound the third and fourth terms in (\ref{eq:Ualph4}) which leads to
\begin{align}
u(\alpha',\sigma'^2)  &\geq  2\ln2\left[\mathcal{C}(\sigma'^2)-\alpha'\right] +\ln\frac{x^{(3)}}{1+\sigma'^2} +  \ln \frac{\sigma'^2}{\sigma'^2(1+e^{-\frac{1}{\sigma}})} 
 -  1  + \frac{1+\sigma'^2}{x^{(3)}}- \frac{1}{\alpha'( x^{(3)})^2}\\
&=  2\ln2\left[\mathcal{C}(\sigma'^2)-\alpha'\right] +\ln\frac{x^{(3)}}{1+\sigma'^2} - \ln (1+e^{-\frac{1}{\sigma}}) 
 -  1  + \frac{1+\sigma'^2}{x^{(3)}}- \frac{1}{\alpha'( x^{(3)})^2}
\label{eq:Ualph5}
\end{align}
Applying a simple lower bound on the second term on the right hand side of (\ref{eq:Ualph5}) implies
\begin{align}
u(\alpha',\sigma'^2)  &\geq  2\ln2\left[\mathcal{C}(\sigma'^2)-\alpha'\right]  +  1  - \frac{1+\sigma'^2}{x^{(3)}} - \ln (1+e^{-\frac{1}{\sigma}}) 
 -  1  + \frac{1+\sigma'^2}{x^{(3)}}- \frac{1}{\alpha'( x^{(3)})^2}\\ 
 &= 2\ln2\left[\mathcal{C}(\sigma'^2)-\alpha'\right]   - \ln (1+e^{-\frac{1}{\sigma}}) 
- \frac{1}{\alpha'( x^{(3)})^2}\ .
\label{eq:Ualph6}
\end{align}
Now we notice that the fixed point $x^{(3)}$ of (\ref{eq:char}) is monotonically increasing both as a function of $\alpha'$ and $\sigma'^2$ and, therefore,
$x^{(3)}(\alpha',\sigma'^2) \geq x^{(3)}(4,0) = 0.7396$ for $\alpha'\geq 4$, $\sigma'^2\geq 0$. Hence
\begin{align}
u(\alpha',\sigma'^2)  &\geq 2\ln2\left[\mathcal{C}(\sigma'^2)-\alpha'\right]   - \ln (1+e^{-\frac{1}{\sigma}}) 
- \frac{1}{\alpha'( x^{(3)}(4,0))^2} \\
&= 2\ln2\left[\mathcal{C}(\sigma'^2)-\alpha'\right]   - \ln (1+e^{-\frac{1}{\sigma}}) 
- \frac{1.8281}{\alpha'}\ .
\label{eq:Ualph7}
\end{align}
which simplifies to is a quadratic equation with respect to $\alpha'$. A candidate solution 
\begin{equation}
\alpha'=\alpha^{**} = \mathcal{C}(\sigma'^2)  - \frac{1.08}{\mathcal{C}(\sigma'^2) } 
\label{eq:alphastarstar}
\end{equation}
guarantees $u(\alpha',\sigma'^2) > 0$ for $\sigma'^2$ such that $\alpha^{**}\geq 4$ ($\mathcal{C}(\sigma^2) \geq 4.26$).
We can now bound the gap to capacity based on parameters $\alpha^{**},\sigma'^2$ 
\begin{align}
\mathcal{C}(\sigma'^2) &- \alpha^{**}  \mathcal{C}\stxt{BIAWGN} \left(\frac{1}{\alpha^{**} x^{(1)}}\right) \leq  \mathcal{C}(\sigma'^2) \left[1- \mathcal{C}\stxt{BIAWGN} \left(\frac{1}{\alpha^{**} x^{(1)}}\right) \right] + \frac{1.08}{\mathcal{C}(\sigma'^2)}\\
 &\leq  2\ln2 \mathcal{C}(\sigma'^2) \int_{\frac{1}{\alpha' x^{(1)}}}^\infty g_2(\gamma) d \gamma  + \frac{1.08}{\mathcal{C}(\sigma'^2)} \leq   2\ln2 \mathcal{C}(\sigma'^2) \int_{\frac{1}{\alpha^{**} x^{(1)}}}^\infty \pi Q(\sqrt{\gamma}) d \gamma  + \frac{1.08}{\mathcal{C}(\sigma'^2)}\\
& \leq 2\ln2 \mathcal{C}(\sigma'^2) 2\pi \exp\left(-\frac{1}{2x^{(1)}}\right)  + \frac{1.08}{\mathcal{C}(\sigma'^2)} \leq 2\ln2 \mathcal{C}(\sigma'^2) 2\pi \exp\left(-\frac{1}{4\sigma'^2}\right)  + \frac{1.08}{\mathcal{C}(\sigma'^2)}  \leq  \frac{1.081}{\mathcal{C}(\sigma^2)}
\end{align}
where we used an upper bound  
on the MSE function from~\cite{BurSchShiKry2004}, 
the MSE-capacity relationship~\cite{guover05}, 
and then an upper bound $Q(\sqrt{\gamma}) \leq -\exp(\gamma/2)$. 

We then bound the gap between the actual capacity 
\[
\mathcal{C}(\sigma^2) - \mathcal{C}(\sigma'^2) = \frac{1}{2}\log_2\left(\frac{\sigma^2+1}{\sigma^2+1-\frac{1}{L}}\right) 
\]
and the achievable rate 
\begin{align}
&\alpha^{**}  \mathcal{C}\stxt{BIAWGN} \left(\frac{1}{\alpha^{**} x^{(1)}}\right) -\mathcal{R}\stxt{coup}(\alpha^{**},\sigma^2) \leq \alpha^{**}  \mathcal{C}\stxt{BIAWGN} \left(\frac{1}{\alpha^{**} x^{(1)}}\right) - \alpha  \mathcal{C}\stxt{BIAWGN} \left(\frac{1}{\alpha^{**} x^{(1)}}\right) \\
&\leq \alpha^{**} - \alpha = \frac{L-1}{M-1}- \frac{L}{M} \leq \frac{\alpha-1}{M-1} \leq \frac{\mathcal{C}(\sigma^2)-1}{M-1}
\end{align}


The Theorem is proved.
\end{proof}

\subsection{Proof of Theorem~\ref{thm:scaling}}
\label{sec:scaling}

Consider $(\alpha,\sigma^2)$ satisfying the conditions of Theorem~\ref{thm:mainthm} with $\sigma^2$ chosen such that $\mathcal{C}(\sigma^2) >4.26$ and 
\begin{equation}
\alpha \in [\mathcal{C}(\sigma^2)-1.081/\mathcal{C}(\sigma^2),\mathcal{C}(\sigma^2)].
\label{eq:alphainterval}
\end{equation}
We apply the low-SNR MSE $g\stxt{mse}(\eta)$ series expansion~\cite{verdu11}  for $\eta \rightarrow 0$
\begin{equation}
g\stxt{mse}(\eta) = 1 -\eta + \eta^2 +  \mathcal{O}(\eta^3)
\label{eq:mseser}
\end{equation}
which is equivalent to the fact that there exists an $\eta_1 >0$ and a constant $C_1 > 0$ such that for $\eta \in [ 0, \eta_1]$
\begin{equation}
g\stxt{mse}(\eta) = 1 -\eta + \eta^2 +  f(\eta) \quad \quad \textrm{where} \quad \quad |f(\eta)| \leq C_1 \eta^3.
\label{eq:gmseser}
\end{equation}
Consider now the third root $x^{(3)}$ of the characteristic equation   
\begin{equation}
x^{(3)} = g\stxt{mse}\left (\frac{1}{\alpha x^{(3)}}\right) + \sigma^2
\label{eq:x3char}
\end{equation}
that, according to Lemma~\ref{lem:roots} (c) satisfies
\begin{align}
1- \frac{1}{\alpha} - \frac{2}{\alpha^2}   \leq 1+\sigma^2 - \frac{1}{\alpha(1+\sigma^2)} - \frac{2}{\alpha^2(1+\sigma^2)^3}  \leq x^{(3)} \leq 1+\sigma^2\ .
\label{eq:x3lowerb}
\end{align}
The lower bound (\ref{eq:x3lowerb}) implies that $1/(\alpha x^{(3)}) \rightarrow 0$ for $\sigma^2 \rightarrow 0$ and $\alpha$ satisfying (\ref{eq:alphainterval}). Therefore, the series expansion (\ref{eq:gmseser}) can be applied to (\ref{eq:x3char}). There exists a $\sigma_1$ such that $\sigma < \sigma_1$ implies $1/(\alpha x^{(3)}) < \eta_1$ and, therefore,
\begin{equation}
x^{(3)} =  1 -\frac{1}{\alpha x^{(3)}}+ \frac{1}{(\alpha x^{(3)})^2} +  f\left(\frac{1}{\alpha x^{(3)}}\right) \quad \quad \textrm{where} \quad \quad \left|f\left(\frac{1}{\alpha x^{(3)}}\right)\right| \leq C_1 \frac{1}{(\alpha x^{(3)})^3}.
\label{eq:x3ser1}
\end{equation}
\begin{proposition}
There exists $\sigma_2$ and a constant $C_2>0$ such that for $\sigma < \sigma_2$ and $\alpha$ satisfying (\ref{eq:alphainterval})
\begin{equation}
x^{(3)} =  1 -\frac{1}{\alpha } +  f_2\left(\frac{1}{\alpha}\right) \quad \quad \textrm{where} \quad \quad \left|f_2\left(\frac{1}{\alpha}\right)\right| \leq C_2 \frac{1}{\alpha^3}.
\label{eq:x3ser2}
\end{equation}
or equivalently
\begin{equation}
x^{(3)} =  1 -\frac{1}{\alpha } + \mathcal{O}\left(\frac{1}{\alpha^3}\right)
\label{eq:x3ser3}
\end{equation}
as $\sigma \rightarrow 0$ and $\alpha \rightarrow \infty$ ($\alpha$ satisfies (\ref{eq:alphainterval})).
\label{prop:x3exp}
\end{proposition}
\begin{proof}
First of all (\ref{eq:x3lowerb}) directly implies that there exists a $\sigma_2$  such that for  $\sigma < \sigma_2$ and $\alpha$ satisfying (\ref{eq:alphainterval})
\[
\left| x^{(3)} - \left(1 - \frac{1}{\alpha}\right)\right| \leq \frac{3}{\alpha^2}\ .
\]
Consider now (\ref{eq:x3ser1}) and subtract the first two terms on th right hand side of (\ref{eq:x3ser2}) from it
\begin{align}
\left| x^{(3)}  - \left(1 -\frac{1}{\alpha }\right) \right| &= \left| 1 -\frac{1}{\alpha x^{(3)}}+ \frac{1}{(\alpha x^{(3)})^2} +  f\left(\frac{1}{\alpha x^{(3)}}\right)  - 1 +\frac{1}{\alpha }+ \frac{1}{\alpha^2}- \frac{1}{\alpha^2}\right|\\
&= \left|\frac{1}{\alpha}\left(1-\frac{1}{ x^{(3)}} + \frac{1}{ \alpha }  \right) - \frac{1}{\alpha^2}\left(1-\frac{1}{(x^{(3)})^2}\right)  +  f\left(\frac{1}{\alpha x^{(3)}}\right)  \right| \\
&\leq \frac{1}{\alpha^2x^{(3)}}\left|\alpha x^{(3)} - \alpha + x^{(3)}\right| + \frac{1}{\alpha^2}\left|1-\frac{1}{(x^{(3)})^2}\right| +  \frac{C_1}{\alpha^3}\\
&\leq \frac{2}{\alpha^2}\left|\alpha x^{(3)} - \alpha + x^{(3)}\right| + \frac{1}{\alpha^2}\left|x^{(3)}-1\right|\frac{1+x^{(3)}}{(x^{(3)})^2} +  \frac{C_1}{\alpha^3}\\
&\leq \frac{2}{\alpha^2}\left(\frac{1}{\alpha}+\frac{6}{\alpha^2}\right) + \frac{1}{\alpha^2}\left(\frac{1}{\alpha}+\frac{3}{\alpha^2}\right)12 +  \frac{C_1}{\alpha^3} \leq \frac{C_2}{\alpha^3}
\end{align}
where $C_2$ is a constant. Proposition is proved.
\end{proof}
Proposition~1 implies
\[
\ln \left( x^{(3)}\right) = -\frac{1}{\alpha} - \frac{1}{2}\frac{1}{\alpha^2} + \mathcal{O}\left(\frac{1}{\alpha^3}\right)
\]
as $\sigma \rightarrow 0$ and $\alpha \rightarrow \infty$ ($\alpha$ satisfies (\ref{eq:alphainterval})). Based on the relationship between the MSE (\ref{eq:mseser}) and the channel capacity~\cite{verdu11} we have
\begin{align}
\mathcal{C}_\mathcal{A}\left(\eta\right)  = \frac{1}{2 \ln 2}\left( \eta - \frac{\eta^2}{2} +  \frac{\eta^3}{3} \right) + \mathcal{O}(\eta^3)
\label{eq:CA}
\end{align}
for $\eta \rightarrow 0$ where the fraction $1/(2\ln 2)$ is applied since the information we consider is measured in bits. Hence, using the above capacity expression and Proposition~\label{prop:x3exp} for $\eta = 1/(\alpha x^{(3)})$ we obtain
\begin{align}
2 \ln 2 \alpha \mathcal{C}_\mathcal{A}\left(\eta\right)  &= \frac{1}{x^{(3)}} - \frac{1}{2\alpha \left(x^{(3)}\right)^2} +  \frac{1}{3\alpha^2 \left(x^{(3)}\right)^3} +  \mathcal{O}\left(\frac{1}{\alpha^3}\right)\\
&= 1 + \frac{1}{\alpha} + \frac{1}{\alpha^2}  - \frac{1}{2\alpha} \left( 1 + \frac{2}{\alpha} \right) +  \frac{1}{3\alpha^2 } +  \mathcal{O}\left(\frac{1}{\alpha^3}\right) =1 + \frac{1}{2\alpha} + \frac{1}{3\alpha^2}  +  \mathcal{O}\left(\frac{1}{\alpha^3}\right)
\end{align}
as $\sigma \rightarrow 0$ and $\alpha \rightarrow \infty$ ($\alpha$ satisfies (\ref{eq:alphainterval})). Coming back to the potential function (\ref{eq:Ualph2}) we compute the asymptotics of the potential function in terms of $\alpha$
\begin{align}
u(\alpha,\sigma^2) &= \ln \frac{x^{(3)}}{x^{(1)}} - \frac{\sigma^2}{x^{(1)}} + \frac{\sigma^2}{x^{(3)}} - 2 \ln 2 \alpha \mathcal{C}_\mathcal{A}\left(\frac{1}{\alpha x^{(1)}}\right) + 2 \ln 2 \alpha \mathcal{C}_\mathcal{A}\left(\frac{1}{\alpha x^{(3)}}\right)\\
&= \ln \sigma^{-2} + \ln x^{(3)} - 1 - 2 \ln 2 \alpha + 2 \ln 2 \alpha \mathcal{C}_\mathcal{A}\left(\frac{1}{\alpha x^{(3)}}\right) + \mathcal{O}\left(\frac{1}{\alpha^3}\right)\\
 &= \ln \left(1+\frac{1}{\sigma^2}\right)  - \frac{1}{\alpha} -  \frac{1}{2\alpha^2}  - 1 - 2 \ln 2 \alpha + 1 + \frac{1}{2\alpha} + \frac{1}{3\alpha^2} + \mathcal{O}\left(\frac{1}{\alpha^3}\right)\\
 &= \ln  \left(1+\frac{1}{\sigma^2}\right)   - \frac{1}{2\alpha} -  \frac{1}{6\alpha^2}  - 2 \ln 2 \alpha  + \mathcal{O}\left(\frac{1}{\alpha^3}\right)
\label{eq:pota}
\end{align}
as $\sigma \rightarrow 0$ and $\alpha \rightarrow \infty$ ($\alpha$ satisfies (\ref{eq:alphainterval})). By selecting 
\[
\alpha =  \mathcal{C}(\sigma^2) -  \frac{1}{4 \ln 2 \mathcal{C}(\sigma^2) } + \frac{1}{12 \ln 2 \mathcal{C}(\sigma^2) ^2} + \mathcal{O}\left(\frac{1}{\mathcal{C}(\sigma^2)^3}\right)
\]
we ensure that the potential function  is positive and the iterative decoding converges to the smallest root $x^{(1)}$ of the characteristic equation. The resulting gap to capacity then scales as 
\begin{align}
\mathcal{C}(\sigma^2) - \alpha \mathcal{C}_\mathcal{A}\left(\frac{1}{\alpha x^{(1)}}\right)  =  \frac{1}{4 \ln 2 \mathcal{C}(\sigma^2) } + \frac{1}{12 \ln 2 \mathcal{C}(\sigma^2) ^2} + \mathcal{O}\left(\frac{1}{\mathcal{C}(\sigma^2)^3}\right)
\end{align}
for $\sigma\rightarrow 0$. The Theorem is proved.


%
%

\end{document}